\documentclass[aps, prd, twocolumn, reprint, superscriptaddress, nofootinbib,eqsecnum]{revtex4}
\usepackage{amsmath}
\usepackage{amssymb}
\usepackage{mathtools}
\usepackage{bm}
\usepackage{hyperref}
\usepackage{units}
\usepackage{graphicx}
\usepackage[usenames,dvipsnames]{xcolor}
\usepackage{soul}
\usepackage{color}
\usepackage{quantumDefinitions}

\newcommand\parsymb{\text{{\tiny ((}}}
\newcommand\antiparsymb{\text{{\tiny )(}}}
\newcommand\rpart{\text{r}}
\newcommand\ipart{\text{i}}

\renewcommand{\vec}[1]{\bm{#1}}

\renewcommand{\Re}{\operatorname{Re}}
\renewcommand{\Im}{\operatorname{Im}}
\DeclareMathOperator{\csch}{csch}

\DeclareMathOperator{\arccosh}{arccosh}
\DeclareMathOperator{\res}{Res}

\makeatletter
\@ifundefined{textcolor}{}
{%
 \definecolor{BLACK}{gray}{0}
 \definecolor{WHITE}{gray}{1}
 \definecolor{RED}{rgb}{1,0,0}
 \definecolor{GREEN}{rgb}{0,.4,0}
 \definecolor{BLUE}{rgb}{0,0,1}
 \definecolor{CYAN}{cmyk}{1,0,0,0}
 \definecolor{MAGENTA}{cmyk}{0,1,0,0}
 \definecolor{YELLOW}{cmyk}{.2,.4,1,0}
 }
\makeatother

\newcommand{\blk}{\protect\color{BLACK}}

\newcommand{\abs}[1]{\left\lvert{#1}\right\rvert}

\begin{document}
\title{Acceleration-assisted entanglement harvesting and rangefinding}

\date{\today}

\author{Grant Salton}
\affiliation{Stanford Institute for Theoretical Physics, Stanford University, Stanford, CA 94305}
\affiliation{Department of Physics, McGill University, 3600 rue University, Montreal, Quebec, H3A 2T8, Canada}
\affiliation{Department of Physics \& Astronomy, University of Waterloo, 200 University Ave W., Waterloo, Ontario, N2L 3G1, Canada}
\author{Robert B. Mann}
\affiliation{Department of Physics \& Astronomy, University of Waterloo, 200 University Ave W., Waterloo, Ontario, N2L 3G1, Canada}
\affiliation{Institute for Quantum Computing, University of Waterloo, 200 University Ave W., Waterloo, Ontario, N2L 3G1, Canada}
\author{Nicolas C. Menicucci}
\affiliation{School of Physics, The University of Sydney, Sydney, NSW 2006, Australia}
\affiliation{Perimeter Institute for Theoretical Physics, 31 Caroline Street, Waterloo, Ontario, N2L 2Y5, Canada}

\begin{abstract}
We study entanglement harvested from a quantum field through local interaction with Unruh-DeWitt detectors undergoing linear acceleration.  The interactions allow entanglement to be swapped locally from the field to the detectors. We find an enhancement in the entanglement harvesting by two detectors with anti-parallel acceleration over those with inertial motion.  This enhancement is characterized by the presence of entanglement between two detectors that would otherwise maintain a separable state in the absence of relativistic motion (with the same distance of closest approach in both cases).  We also find that entanglement harvesting is degraded for two detectors undergoing parallel acceleration in the same way as for two static, comoving detectors in a de~Sitter universe.  This degradation is known to be different from that of two inertial detectors in a thermal bath.  We comment on the physical origin of the harvested entanglement and present three methods for determining distance between two detectors using properties of the harvested entanglement.  Information about the separation is stored nonlocally in the joint state of the accelerated detectors after the interaction; a single detector alone contains none. We also find an example of entanglement sudden death exhibited in parameter space.
\end{abstract}

\maketitle


\section{Introduction}
\label{sec:intro}

The ground state of a system of coupled harmonic oscillators is the unique state with zero excitations in any normal mode of the system. With respect to these \emph{global} modes, the ground state is unentangled (separable)~\cite{Horodecki2009}. We can equally well describe the same state in the tensor-product basis of number states of each individual oscillator. With respect to these \emph{local} modes, the ground state of the system is entangled (unless the coupling is everywhere zero). This property is generic: the same state may be considered either separable or entangled depending on the tensor-product structure we choose for the state space, with operational locality often determining the proper choice if we wish to use the entanglement as a physical resource~\cite{Zanardi2004}.

It is in this sense that we say that the vacuum of a free quantum field in Minkowski spacetime is entangled with respect to the state space of local observables~\cite{Summers1987}, even though it is separable with respect to Minkowski plane-wave modes, with zero particles in every mode. This fact is borne out operationally by the ability to swap this entanglement to local quantum systems (``detectors'') through local interactions~\cite{Reznik2005}.  Local hidden-variable  models cannot account for the long-distance correlations due to this entanglement \cite{Reznik2005,Verch:2004vj}, which decrease at a rate slower than $\exp[-(L/cT )^3]$, where $L$ is the separation and $T$ the duration of the detector-field coupling.  Other types of vacuum correlations, such as true multi-region entanglement \cite{Acin2001} and full nonlocality \cite{PhysRevD.35.3066,PhysRevLett.88.170405}, are also possible.  Time-like separated detectors can also swap entanglement from the vacuum of a massless field \cite{Olson:2010jy}, even though the field has  lightlike excitations. Superconducting circuits or quantum optical settings may soon admit experimental  realizations of these phenomena  \cite{PhysRevLett.107.150402,PhysRevLett.109.033602}. 

Since entanglement is a phenomenon that is uniquely quantum mechanical in nature~\cite{Horodecki2009,Masanes:2008ib} and can be considered both an information-theoretic and a physical resource~\cite{Blume-Kohout2002}, we call this swapping process \emph{entanglement harvesting} to connote the reaping of a naturally-occurring and useful resource from a quantum field.\footnote{The term `entanglement harvesting' was coined by one of us~(N.C.M.)\ in early 2012. Its first public use was in a talk by another of us~(G.S.)\ at the Relativistic Quantum Information Conference at Perimeter Institute in June 2012, which reported on the initial results of this project. Its first use in print is in a 2012 review article~\cite{MartinMartinez:2012gh} by Eduardo Mart\'in-Mart\'inez and N.C.M. 
} Just as factors like sunlight, water, and nutrients affect the crops harvested from a field on a farm, entanglement harvested from a quantum field is affected by spacetime curvature~\cite{VerSteeg2009}, temperature~\cite{Braun2005,VerSteeg2009}, and the type of field used~\cite{Nambu2011,Nambu2013}. Also, just as in the farming case, the tools used for the harvesting---in this case, the detectors' state of motion~\cite{Massar2006} and local coupling~\cite{Reznik2005}---also affect the amount and type of entanglement obtained. This harvesting can also be repeated, resulting in what is now known as \emph{entanglement farming}\blk~\cite{Martin-Martinez:2013eda}, in which successive pairs of unentangled particles sent through a suitably prepared cavity will reliably emerge significantly entangled.

The phenomenon of entanglement harvesting will necessarily involve a combination of relativistic and quantum effects. Conceptually important qualitative differences emerge when relativistic effects are included in studies of quantum information~\cite{Peres2004}. For example, entanglement was found to be an observer-dependent property that is degraded from the perspective of accelerated observers moving in flat spacetime~\cite{FuentesSchuller2005,Alsing2006,Leon2009,Mann2009}.  Entanglement between modes of either bosonic or fermionic fields are degraded from the perspective of observers moving in uniform acceleration. These effects are present in both black hole spacetimes~\cite{Ahn2008} and in cosmological scenarios~\cite{VerSteeg2009,Ball2006,Fuentes2010,MartinMartinez:2012gh}.

A common approach employed in many of these studies is the single-mode approximation.  This approximation attempts to relate a single-frequency Minkowski mode (observed by inertial observers) with a single  frequency Rindler mode (observed by uniformly accelerated observers).  This approximation does not hold for general states but only for a specific kind of Unruh mode~\cite{Bruschi2010}.

Here we study entanglement between two hypothetical, uniformly accelerated, quantum particle detectors.  We use the terms ``detectors'' and ``observers'' interchangeably.  The detectors live in (3+1)-dimensional Minkowski spacetime, within which exists a quantum field in the Minkowski vacuum state~$\ket{0}$.  We work with a massless scalar field for simplicity.  The detectors' accelerated trajectories give rise to Unruh radiation~\citep{Davies1975,Unruh1976} with a well known thermal profile corresponding to a temperature
\begin{equation}
T=\frac{\kappa}{2\pi k_{B}},
\label{eq:tempacc}
\end{equation}
where $\kappa$ is the magnitude of the acceleration, $k_{B}$ is Boltzmann's constant, and we work in natural units with $c=\hbar=1$.  We find that  for two detectors undergoing \emph{parallel} acceleration, entanglement harvesting is degraded in exactly the same way as for two static, comoving detectors in a de Sitter universe (with a conformally coupled field~\cite{Birrell1982}), but distinct from that of two inertial detectors in a thermal bath, as previously shown in Ref.~\cite{VerSteeg2009}. Somewhat more remarkably, we  find an enhancement of entanglement harvesting by two detectors with \emph{anti-parallel} acceleration,  characterized by the presence of entanglement between two detectors that would otherwise maintain a separable state in the absence of relativistic motion.

\section{Analysis}
\label{sec:analysis}

\subsection{Detector-field interaction}

The details of the setup closely follow Ref.~\citep{VerSteeg2009} with any modifications clearly described. The background field is a massless scalar field in ${(3+1)}$-dimensional Minkowski spacetime~\cite{Birrell1982}, initially in the Minkowski vacuum state. 
When comparing to the results of an expanding universe~\cite{VerSteeg2009}, we will assume conformal coupling~\cite{Birrell1982} because minimal coupling has a very different effect~\cite{Nambu2011,Nambu2013}, but both are identical for Minkowski spacetime (with Ricci scalar $R=0$). We consider two local detectors (defined below), labeled $a$ and $b$, on the following spacetime trajectories:
\begin{align}\label{eq:trajectoriesSymm}\nonumber
x_a^\mu & = (t_a,\vec x_{a}), & x_b^\mu & = (t_b,\vec x_{b}),\\
t_a & = t_a(\tau), &  t_b & = t_b(\tau^\prime), \\\nonumber
\vec x_a & = \vec x_a(\tau), &  \vec x_b & = \vec x_b(\tau^\prime),
\end{align}
where $b$ claims the prime, $\tau$ and $\tau^\prime$ are the proper times for observers $a$ and $b$, respectively, and $\vec x_{i}$ is the spatial trajectory of observer $i$.  The 4-vector index $\mu$ is henceforth dropped.

We model the detectors as point-like, two-level quantum systems with an energy gap~$\Omega$, each of which interacts with the field via an Unruh-DeWitt interaction~\cite{DeWitt:1979tl,Unruh1976}, whose interaction-picture Hamiltonian is
\begin{align}
\label{eq:UdW}
	\hat H_\text{I}(\tau) = \eta(\tau) \hat\phi[x(\tau)] \hat\sigma_x(\tau).
\end{align}
Here $\eta(\tau)$ is a window function (i.e., time-dependent coupling strength) that governs the interaction with the ambient scalar field, $x(\tau)$ is either of the 4-vectors $x_a$ or $x_b$, $\hat\phi[x(\tau)]$ is the interaction-picture field amplitude at spacetime point~$x$, and $\hat\sigma_x(\tau)$ is the interaction-picture Pauli-$x$ operator for the two-level system. The detectors are identical aside from their trajectories, and we formally delay readout (i.e., projective measurement) of the detectors indefinitely because we wish to analyze the quantum features of the resulting state of the detectors. The detectors start in their respective ground states ($\ket g \otimes \ket g$), so if the interaction events are sufficiently far separated, then any entanglement appearing in the final state of the detectors must have been harvested from the field through the local interactions.

We choose $\eta$ at all times to be very small (i.e., $0 < \eta(\tau) \ll 1$), with nontrivial support only over a small range in $\tau$ (as compared to the light-crossing time between them).  This choice gives a short, weak interaction with the field. Its shortness enforces spacelike separation of the detection events, and its weakness allows us to treat these interactions perturbatively.

For calculational simplicity, we take $\eta(\tau)$ to be a Gaussian with variance~$\sigma^2$:
\begin{align}
\label{eq:etadef}
	\eta(\tau) = \eta_0 \exp\left( - \frac {\tau^2} {2\sigma^2} \right),
\end{align}
with $\sigma \ll L$ and $0 < \eta_0 \ll 1$. Note that this function is analytic and therefore cannot be strictly compact in~$\tau$. This means that, in general, the detection events are only approximately spacelike separated. Analyticity of the Gaussian will allow us to evaluate integrals using complex analysis, but the fact that it has vanishing but nonzero support for $\tau \to \pm \infty$ will have interesting consequences that we will explore later on.

\subsection{Single detector response}

The fact that the field is in the (Minkowski) vacuum does not necessarily imply that a detector will fail to become excited when interacting with it~\cite{MartinMartinez:2012gh,Birrell1982}. This is due to counter-rotating terms (such as $\hat\sigma_- \hat a$ and $\hat\sigma_+ \hat a^\dag$) in the expansion of Eq.~\eqref{eq:UdW}. In fact, most window functions and trajectories will produce a nonzero response (excitation probability) in an Unruh-DeWitt detector due to the presence of these terms. Their contribution---and thus the probability of excitation---will tend to 0, however, when the energy gap and coupling strength (during detection) are constant, the trajectory is inertial, and the detection time tends to infinity~\cite{Birrell1982}. 
\blk
%

In general, then, each detector (initially in the ground state) is prone to excitation through its traveling interaction with the field.  To lowest nontrivial order (i.e.,~second order) in the detector coupling strength~$\eta_0$, the probability~$A$ that exactly one of the detectors will be excited is given by~\cite{Birrell1982}
\begin{equation}
A=\int_{-\infty}^{\infty}d\tau\int_{-\infty}^{\infty}d\tau'\eta(\tau)\eta(\tau')e^{-i\Omega(\tau-\tau')}D^+(x_a(\tau);x_a(\tau'))
\label{eq:A}
\end{equation}
where $\Omega$~is the detector energy gap, and $D^+(x;x^\prime) = \langle\hat\phi(x)\hat\phi(x^\prime)\rangle$ is the Wightman function for the field $\hat\phi$, which, for the case of the Minkowski vacuum, is given by Eq.~(3.59) of Ref.~\cite{Birrell1982}:
\begin{equation}
\label{eq:generalDPlus}
	D^+(x_a(\tau);x_a'(\tau'))=\frac{-1}{4\pi^{2}[(t-t'-i\epsilon)^{2}-|\vec x_a-\vec x_a'|^{2}]}.
\end{equation}
Note that, due to symmetry, we could equally well make the trajectory substitution $x_a(\cdot) \mapsto x_b(\cdot)$ and get the same value for~$A$. Also note that all quantities are evaluated with $\epsilon \to 0^+$ at the end.

\subsection{Quantifying the harvested entanglement}

To quantify the harvested entanglement, we follow the standard procedure first proposed by Reznik~\citep{Reznik2003} and used also by later authors~\cite{Reznik2005,VerSteeg2009,Nambu2011,Nambu2013}: We calculate the elements of the reduced density matrix of the detectors and then use the partial transpose criterion~\cite{Peres1996}, which is necessary and sufficient for two qubits~\cite{Horodecki1996}, to detect the presence or absence of entanglement after interaction with the field.

As in previous work~\cite{Reznik2003,Reznik2005,VerSteeg2009,Nambu2011,Nambu2013}, the partial transpose criterion for our setup reduces to a simple comparison between two quantities. Entanglement exists between the detectors if and only if
\begin{equation}
\abs X>A,
\label{eq:EntCond}
\end{equation}
where $A$ is defined in Eq.~\eqref{eq:A}, and $X$ can be interpreted as an amplitude for (virtual) particle exchange between the detectors (to second order in $\eta_0$):
\begin{align}
\label{eq:Xeq}
	X&=-\int_{-\infty}^{\infty}d\tau\int_{-\infty}^{\tau}d\tau'\eta(\tau)\eta(\tau')e^{i\Omega(\tau+\tau')} \nonumber \\
	&\quad \times [D^+(x_{a}(\tau);x_{b}(\tau'))+D^+(x_{b}(\tau);x_{a}(\tau'))],
\end{align}
where the Wightman functions are taken with respect to the two trajectories $x_{a}$ and $x_{b}$ (cf.\ Eq.~\eqref{eq:A}, which only involves one trajectory at a time), and $\Omega$ is the energy gap of the detector. The amount of entanglement harvested can be quantified by the negativity~\citep{Vidal2002} of the final state of the detectors, which in this case is simply
\begin{align}
\label{eq:negativity}
	N = \max \{\abs X - A,0\}\,.
\end{align}
A two-qubit state has nonzero negativity iff it is entangled~\cite{Horodecki1996}.

Entanglement is therefore seen as a competition between the probability of actual detector excitation~($A$) and the magnitude of an off-diagonal term in the density matrix representing the amplitude for particle exchange~($X$)~\cite{Reznik2005}. When the detection events are entirely spacelike separated, the exchange term can only be interpreted as swapping preexisting entanglement out of the field and into the detectors---rather than simply using the field as a transmission medium by which to exchange real excitations. When the detection events are only approximately spacelike separated, as is the case for some choices of parameters here (since $\eta(\tau)$ is Gaussian), there are additional subtleties, which we address later.

\subsection{Accelerating trajectories and comparison with other scenarios}

The canonical formalism for studying accelerated observers in Minkowski spacetime~\cite{Unruh1976,Birrell1982,Crispino2008} involves defining a Rindler wedge corresponding to an accelerating trajectory and using that wedge to define three additional wedges that separate Minkowski spacetime into left, right, future, and past regions with lightsheet boundaries. A key feature of this partitioning is that all four wedges share a common apex. This suggests a natural set of accelerating trajectories: those hyperbolas that share the bounding lightsheets as asymptotes.

%

\begin{figure}[!tb]
\begin{centering}
\includegraphics[width=.45\columnwidth]{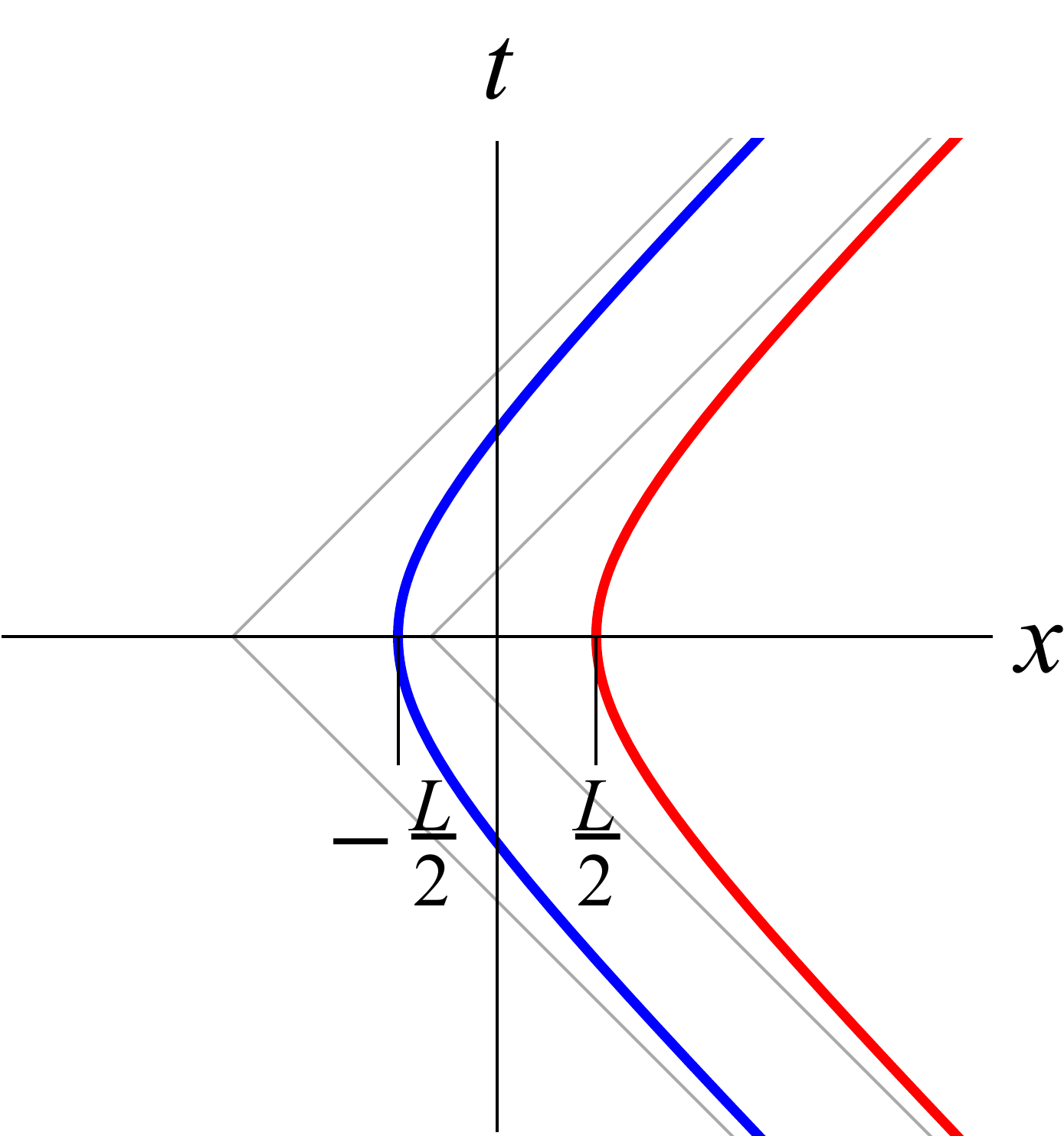}\qquad
\includegraphics[width=.45\columnwidth]{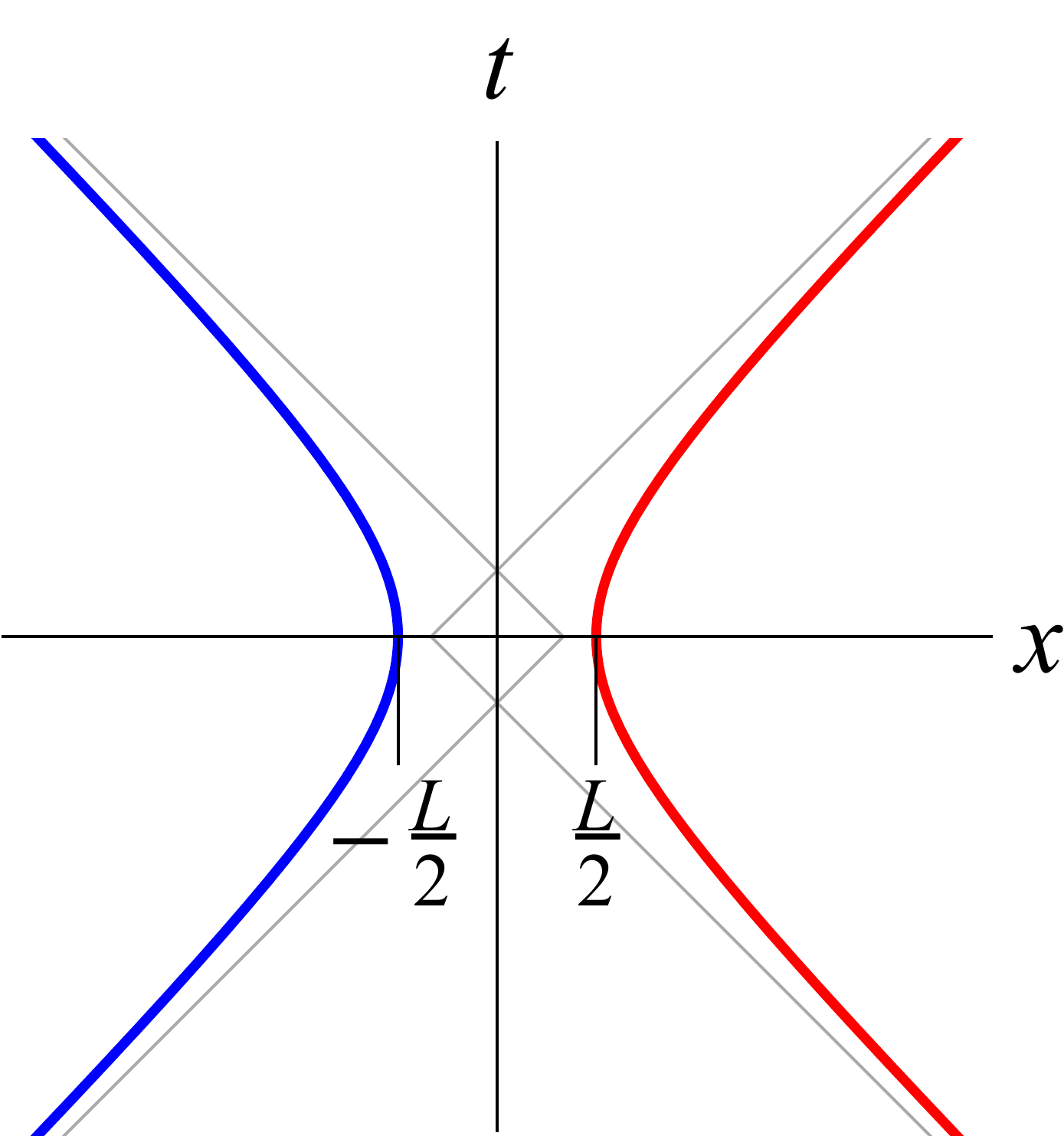}
\end{centering}
\caption[Worldlines of two uniformly accelerated detectors]{\label{fig:Trajectories}(Color online.) Worldlines of two detectors undergoing parallel (left panel) or anti-parallel (right panel) uniform acceleration of equal magnitude. The grey diagonal lines indicate the Rindler wedges associated with each trajectory.  In contrast to the usual treatment of the anti-parallel case~\cite{Birrell1982,Unruh1976,Crispino2008}, the wedges do not necessarily share a common apex. Instead, they could be closer (as shown) or further apart (not shown). The distance of closest approach of the detectors, as measured by an inertial observer at fixed~$x$, is $L$.}
\end{figure}

We take a more operational approach to the problem by focusing on the physical objects: two detectors accelerating uniformly through space. There is no physical reason why the Rindler wedges defined by each observer should be constrained to meet at the origin, and in fact we argue that this approach is restrictive. Forcing the wedges to meet at the origin is tantamount to forcing the distance between the detectors at closest approach, $L$, to be constrained by the acceleration parameter~$\kappa$: specifically, $L=2\kappa^{-1}$. We  relax the condition of wedges with overlapping apexes and study two uniformly accelerated detectors with an arbitrary distance at closest approach.  We also consider both anti-parallel and parallel trajectories, both of which are described by the following trajectories:
\begin{align}\label{eq:acctrajectories}
	x_{a} & = \phantom\pm\frac{1}{\kappa}\Bigl[\cosh(\kappa\tau) - 1\Bigr] +\frac{L}{2}\,, \nonumber \\
	x_{b} & = \pm\frac{1}{\kappa} \Bigl[ \cosh(\kappa\tau) - 1 \Bigr] -\frac{L}{2}\,, \nonumber \\
	 t_a = t_b &= \phantom\pm \frac{1}{\kappa}\sinh(\kappa\tau)\,,
\end{align}
where symmetry allows us to set $\tau' = \tau$ in Eqs.~\eqref{eq:trajectoriesSymm}, the plus/minus sign in~$x_b$ refers to parallel/anti-parallel trajectories, respectively illustrated in left/right panels of Fig.~\ref{fig:Trajectories}, and the other spatial coordinates are taken to be~0.  The quantity $L$ represents the distance of closest approach as measured by an inertial observer along a trajectory of constant $x$. For such an observer, the detectors are separated by this minimum distance when $t=t'=0$.

In what follows, we will use time variables corresponding to the sum and difference of the two proper times:
\begin{align}
\label{eq:xytau}
	x &\coloneqq \tau+\tau'\,, & \tau &= \frac 1 2 (x + y)\,, \nonumber \\
	y &\coloneqq \tau-\tau'\,, & \tau' &= \frac 1 2 (x-y)\,.
\end{align}
Using the trajectories, we define
\begin{align}
\label{eq:Dxydef}
	D(x,y) &\coloneqq \frac 1 2 \Bigl[ D^+\bigl( x_a(\tau); x_b(\tau') \bigr) + D^+\bigl( x_b(\tau); x_a(\tau') \bigr) \Bigr]
	\nonumber \\
	&= D^+\bigl( x_a(\tau); x_b(\tau') \bigr)\,,
\end{align}
where the second line follows from the symmetry of the trajectories---both for Eqs.~\eqref{eq:acctrajectories} and for the ones from Ref.~\cite{VerSteeg2009}, which will be used for comparison. Furthermore,
\begin{align}
\label{eq:Ddetectdef}
	D_{\text{detect}}(y) &\coloneqq D^+(x_a(\tau);x_a'(\tau'))
\end{align}
or equivalently using $x_b(\cdot)$ instead. In fact, for the particular trajectories in question, these formulas are all invariant under the exchange~$x_a(\cdot) \leftrightarrow x_b(\cdot)$.

Using trajectories~\eqref{eq:acctrajectories}, we compute the quantity $X$ from Eq.~\eqref{eq:Xeq}.  We find this integral to be very difficult to compute analytically and to suffer from convergence issues when evaluated using naive numerical methods.  Fortunately, we can calculate the integral by completing the square in the exponential and shifting the contour of integration to convert the complex Gaussian into a real Gaussian.  By making these transformations, we remove the highly oscillatory prefactor and allow for evaluation of the integral using Cauchy's residue theorem.  In particular, we obtain 
\begin{align}
\label{eq:shiftedX}
	e^{(\sigma\Omega)^2}X &= -\frac{\eta_0^2}{2}\int _{-\infty }^{\infty } dx \int _{0}^{\infty} dy\, e^{-\frac{y^2+x^2}{4\sigma^2}}D(x+2i\sigma^2\Omega,y) \nonumber \\
	&\quad+(\text{residue terms}),
\end{align}
The inverse-Gaussian prefactor comes from completing the square, and we have shifted the contour of integration up by $2i\sigma^2\Omega$ and changed variables (see Appendix~\ref{sec:ResidueCalculation}).  Shifting the contour of integration will, in general, cross poles, which give rise to the residue contributions noted explicitly in Eq.~\eqref{eq:shiftedX}.  The pole locations are functions of the input parameters $L$ and $\Omega$, as well as functions of $x$ and $y$.  As such, computing the residue contributions proves challenging, and details are presented in Appendix~\ref{sec:ResidueCalculation}.

We will evaluate $X$ in several different scenarios. The ones that are the focus of this work are parallel and anti-parallel acceleration, as shown in Fig.~\ref{fig:Trajectories}. For the parallel case we use $D(x,y)=D_\parsymb(x,y)$, with
\begin{align}
\label{eq:Daccpara}
	D_\parsymb &= \frac{\kappa^2}{16\pi^2}
	\left[\frac {L\kappa} {2} + i\epsilon - e^{-\frac{x\kappa}{2}} \sinh \left(\frac{y \kappa}{2}\right) \right]^{-1}
	\nonumber \\ &\quad\quad \times
	\left[\frac {L\kappa} {2} - i\epsilon + e^{\frac{x\kappa}{2}} \sinh \left(\frac{y \kappa}{2}\right) \right]^{-1}.
\end{align}
Similarly, for the anti-parallel case we use $D(x,y)=D_\antiparsymb(x,y)$, with
\begin{align}
\label{eq:Daccanti}
	D_\antiparsymb &= \frac{\kappa^2}{16\pi^2}
	\left[e^{-\frac{y\kappa}{2}} \left(\frac {L\kappa} {2}-1\right) -i\epsilon + \cosh \left(\frac{x \kappa}{2}\right) \right]^{-1}
	\nonumber \\ &\quad\quad \times
	\left[e^{\frac{y\kappa}{2}} \left(\frac {L\kappa} {2} - 1 \right) +i\epsilon + \cosh \left(\frac{x \kappa}{2}\right) \right]^{-1}.
\end{align}
The subscripts on $D_\parsymb$ and $D_\antiparsymb$ are used to visually indicate the respective trajectories in Fig.~\ref{fig:Trajectories}. 

For comparison with known results in the literature~\cite{VerSteeg2009,Nambu2013}, we will also need the following additional Wightman functions. The first is for two comoving detectors in de~Sitter spacetime with expansion rate~$\kappa$, separated by constant comoving distance~$L$~\cite{Birrell1982,Nambu2013}:
\begin{align}
\label{eq:DdS}
	D_\text{dS}=&\frac{\kappa^2}{16\pi^2}\left\{e^{x \kappa } \left( \frac{L\kappa} {2} \right)^2- \sinh^2\left[\frac{\kappa}{2} (y-i\epsilon) \right]\right\}^{-1}.
\end{align}
A second Wightman function is for two inertial detectors with fixed proper distance~$L$ in the case where the field has been heated to a finite temperature  $T=(2\pi k_{B})^{-1} \kappa$ in the rest frame of the detectors~\cite{Weldon2000,Nambu2013}:
\begin{align}
\label{eq:Dth}
	D_\text{th}=&\frac{\kappa}{16\pi^2L}\left\{\coth\left[\frac{\kappa}{2} (L-y+i\epsilon)\right] \right. \nonumber \\
	&\quad\quad\;\; +\left.\coth\left[\frac{\kappa}{2} (L+y-i\epsilon)\right]\right\}. 
\end{align}
Any of the four Wightman functions above can substitute for $D(x,y)$ when calculating~$X$ in Eq.~\eqref{eq:shiftedX}.

We use a similar method to calculate the excitation probability~$A$, although in this case we can avoid crossing any poles by restricting our analysis to $\kappa \sigma^2 \Omega < \pi$ (see Appendix~\ref{sec:ResidueCalculation} for details).  As such, the residue contributions vanish here, giving
\begin{align}
\label{eq:shiftedA}
e^{(\sigma\Omega)^2}A&=\eta_0^2\sqrt{\pi}\sigma\int_{-\infty}^{\infty} dy\, e^{-\frac{y^2}{4\sigma^2}} D_{\text{detect}}(y-2i\sigma^2\Omega)
\,.
\end{align}
In this expression, $D_{\text{detect}}(y)$ from Eq.~\eqref{eq:Ddetectdef} is the detector response function for a single detector in a thermal bath at temperature $T=(2\pi k_{B})^{-1} \kappa$~\cite{Birrell1982}:
\begin{align}
\label{eq:DdetectAndDds}
	D_\text{detect}=&\frac{-\kappa^2}{16\pi^2} \csch^2 \left[\frac{\kappa }{2} (y-i\epsilon) \right].
\end{align}

In order to evaluate $A$ and $X$, we employ a saddle point approximation, following Ref.~\cite{Nambu2013}. In particular,
\begin{align}
\label{eq:saddleA}
	A&\simeq  2\pi\eta_0^2\sigma^2e^{-(\sigma\Omega)^2}D_\text{detect}(-2i\sigma^2\Omega)\,, \\
\label{eq:saddleX}
	X&\simeq -2\pi\eta_0^2\sigma^2e^{-(\sigma\Omega)^2}D(2i\sigma^2\Omega,0)
	+ (\text{residue terms}).
\end{align}
Note that there are no residue contributions to $A$ due to the aforementioned restriction $\kappa \sigma^2 \Omega < \pi$. The detector response is identical in all cases considered:
\begin{align}
\label{eq:saddleAeval}
	A &\simeq e^{-(\sigma\Omega)^2} \frac{\eta_0^2} {2 \pi } \left[ \frac {\kappa \sigma} {2} \csc (\kappa  \sigma^2 \Omega) \right]^2\,.
\end{align}
Plugging in for each of the possible choices for~$D$ gives the following for~$X$ in each case:
\begin{align}
\label{eq:saddleXpara}
	X_\parsymb &\simeq -e^{-(\sigma\Omega)^2} \frac{\eta_0^2} {2 \pi } \left( \frac {\sigma} {L} \right)^2 + \text{r.t.}\,, \\
\label{eq:saddleXantipara}
	X_\antiparsymb &\simeq -e^{-(\sigma\Omega)^2} \frac{\eta_0^2} {2 \pi } \left(\frac{\sigma \kappa}{ L\kappa + 2 [\cos \left(\kappa  \sigma^2 \Omega \right)-1]}\right)^2 
	+ \text{r.t.}\,, \\
\label{eq:saddleXdS}
	X_{\text{dS}} &\simeq -e^{-(\sigma\Omega)^2} \frac{\eta_0^2} {2 \pi} \left(\frac{\sigma}{L} \right)^2 e^{-i 2 \kappa  \sigma^2 \Omega}+ \text{r.t.}\,, \\
\label{eq:saddleXth}
	X_{\text{th}} &\simeq -e^{-(\sigma\Omega)^2} \frac{\eta_0^2} {2 \pi} \left( \frac{\kappa \sigma^2} {2L} \right) \coth \left(\frac{L \kappa}{2}\right) + \text{r.t.}\,,
\end{align}
where `r.t.'\ stands for residue terms, which will be evaluated later.

\section{Results: parallel acceleration}
\label{sec:resultspar}

We begin by calculating the negativity~\citep{Vidal2002} for detectors accelerating in the same direction.  Choosing the positive sign in Eqs.~\eqref{eq:acctrajectories} corresponds to two detectors with the desired trajectories, as shown on the left of Fig.~\ref{fig:Trajectories}.  An inertial observer on the trajectory $x(t)=\text{(constant)}$ will measure the detectors to always be separated by a constant distance~$L$. (Note, however, that an observer traveling along with one of the detectors will measure the proper distance to the other detector to be changing over time.) Using the Wightman funtion (\ref{eq:Daccpara}), we cross no poles when shifting the contour of integration in (\ref{eq:shiftedX}), and we are left with only the residue-free contribution.

Plugging $A$ from Eq.~\eqref{eq:saddleAeval} and $X_\parsymb$ from Eq.~\eqref{eq:saddleXpara} into inequality (\ref{eq:EntCond}), we find entanglement whenever
\begin{equation}
\frac {L\kappa} {2} < \sin (\kappa\sigma^2\Omega)\,.
\label{eq:AnalyticParaEnt}
\end{equation}
When the detectors are entangled, the negativity in this case is
\begin{align}
	N_\parsymb &\simeq e^{-(\sigma\Omega)^2} \frac{\eta_0^2} {2 \pi} \left[ \left(\frac {\sigma} {L} \right)^2 - \frac {(\kappa\sigma)^2} {4} \csc^2(\kappa \sigma^2 \Omega) \right]\,.
\end{align}

\subsection{Comparison with expanding and thermal fields}

Interestingly, this negativity formula and entanglement criterion exactly match those for comoving (inertial) observers, at a fixed comoving distance~$L$, in a de~Sitter universe with expansion rate~$\kappa$~\cite{Nambu2013,VerSteeg2009}. Indeed, using $X_\text{dS}$ from Eq.~\eqref{eq:saddleXdS} and verifying that we again cross no poles, we see that $\abs{X_\parsymb} = \abs{X_{\text{dS}}}$, which means that the negativity is the same, $N_{\text{dS}} = N_\parsymb$, and so is the entanglement criterion, Eq.~\eqref{eq:AnalyticParaEnt}. The only differences are in their interpretation: $\kappa$ now refers to the expansion rate instead of the parallel acceleration rate, and $L$ now refers to the comoving distance instead of to the distance as measured by an inertial observer on the trajectory $x(t) = \text{(constant)}$.  Notice that the motion in these two cases seem very different.  In the case of parallel acceleration, the detectors are moving in the same direction, whereas in the de Sitter case the detectors are moving in opposite directions.  However, this intuition is misleading because the similarity between the two is the constancy of $L$.  For parallel acceleration, the distance at closest approach as measured by a stationary observer is constant, while in de Sitter space the detectors are separated by a constant \emph{comoving} distance.

Due to the mathematical connection between Rindler and de~Sitter geometries and the properties of conformal fields (see Chapter 5 of Ref.~\cite{Birrell1982}), one might expect \emph{a~priori} such a connection between the parallel-acceleration case and the case of de~Sitter expansion. We point out a few reasons why this intuition is not enough to prove that they must be the same. First, while $A$ is the same in both cases, $X_\parsymb$ and $X_{\text{dS}}$ actually differ by a phase, Eqs.~\eqref{eq:saddleXpara} and~\eqref{eq:saddleXdS}. As such, while the negativity (which depends only on $\abs X$) is exactly the same in both cases, the actual density matrix in each case is different, resulting in different correlations. Second, the calculations leading to the equivalence involve two approximations---second-order perturbation theory and a saddle-point approximation---that limit the equivalence to the cases where these approximations are valid. Finally, notice that $D_\parsymb$ and $D_{\text{dS}}$ are not the same, Eqs.~\eqref{eq:Daccpara} and~\eqref{eq:DdS}. This leads us to suspect that when the aforementioned approximations break down, the equivalence may be broken as well. On the other hand, our work does not rule out the existence of other coordinates and/or modified trajectories (perhaps parallel acceleration but displaced in a different spatial direction) for which there is exact equivalence between the two cases. This is left as an open problem. \blk

From Ref.~\cite{VerSteeg2009}, we already know that the entanglement profile for two inertial detectors with fixed proper distance~$L$ in a thermal bath in the detectors' rest frame will differ from the de~Sitter case. From the above results, this means it must also differ from the parallel-acceleration case, despite $A$ being the same in all three cases. 
Solving the analogous inequalities using $X_\text{th}$ from Eq.~\eqref{eq:saddleXth} and noting that no poles have been crossed, we find entanglement whenever
\begin{equation}
	\frac {L\kappa} {2} \tanh\left(\frac{L\kappa}{2}\right)< \sin^2(\kappa\sigma^2\Omega),
\label{eq:AnalyticThermalEnt}
\end{equation}
with the temperature~$T$ of the thermal state of the field chosen to be the Gibbons-Hawking temperature associated with $\kappa$~\cite{Gibbons1977}---i.e., $T= (2\pi k_{B})^{-1} \kappa$. Thus, we have reproduced the results first reported in Ref.~\cite{VerSteeg2009} (and later confirmed using this saddle-point method in Ref.~\cite{Nambu2013}), which show that two detectors can distinguish between Gibbons-Hawking and thermal radiation by whether they are able to harvest entanglement or not. If entangled, the negativity in this case is
\begin{align}
	N_{\text{th}} &\simeq e^{-(\sigma\Omega)^2} \frac{\eta_0^2} {2 \pi}
	 \nonumber \\ &\quad \times 
	\left[ \frac{\kappa\sigma^2} {2L} \coth \left(\frac{L \kappa}{2}\right) - \frac {(\kappa\sigma)^2} {4} \csc^2(\kappa \sigma^2 \Omega) \right]\,.
\end{align}

\begin{figure*}[!tb]
\begin{centering}
\includegraphics[width=0.95\columnwidth]{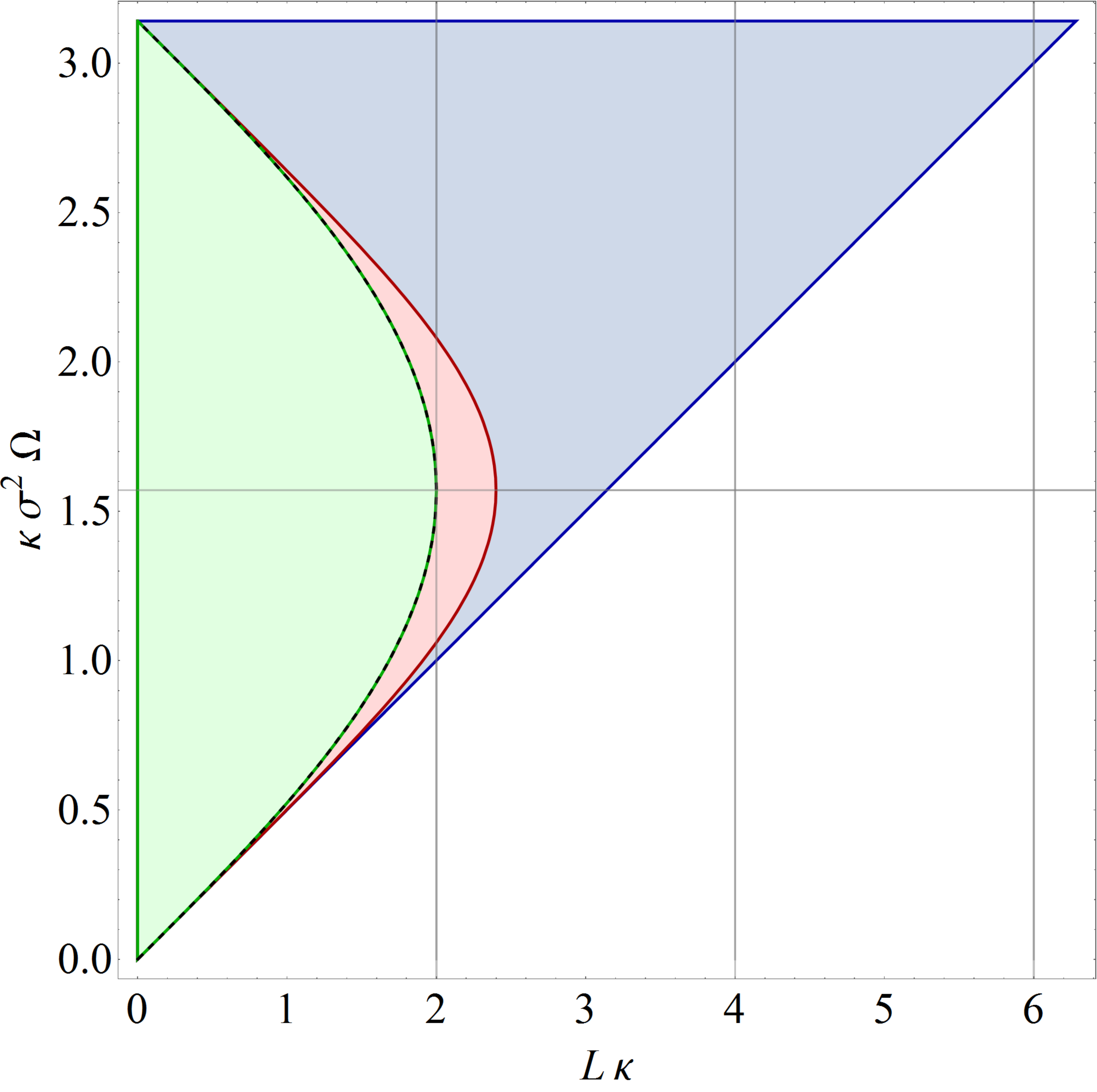}\qquad%
\includegraphics[width=0.95\columnwidth]{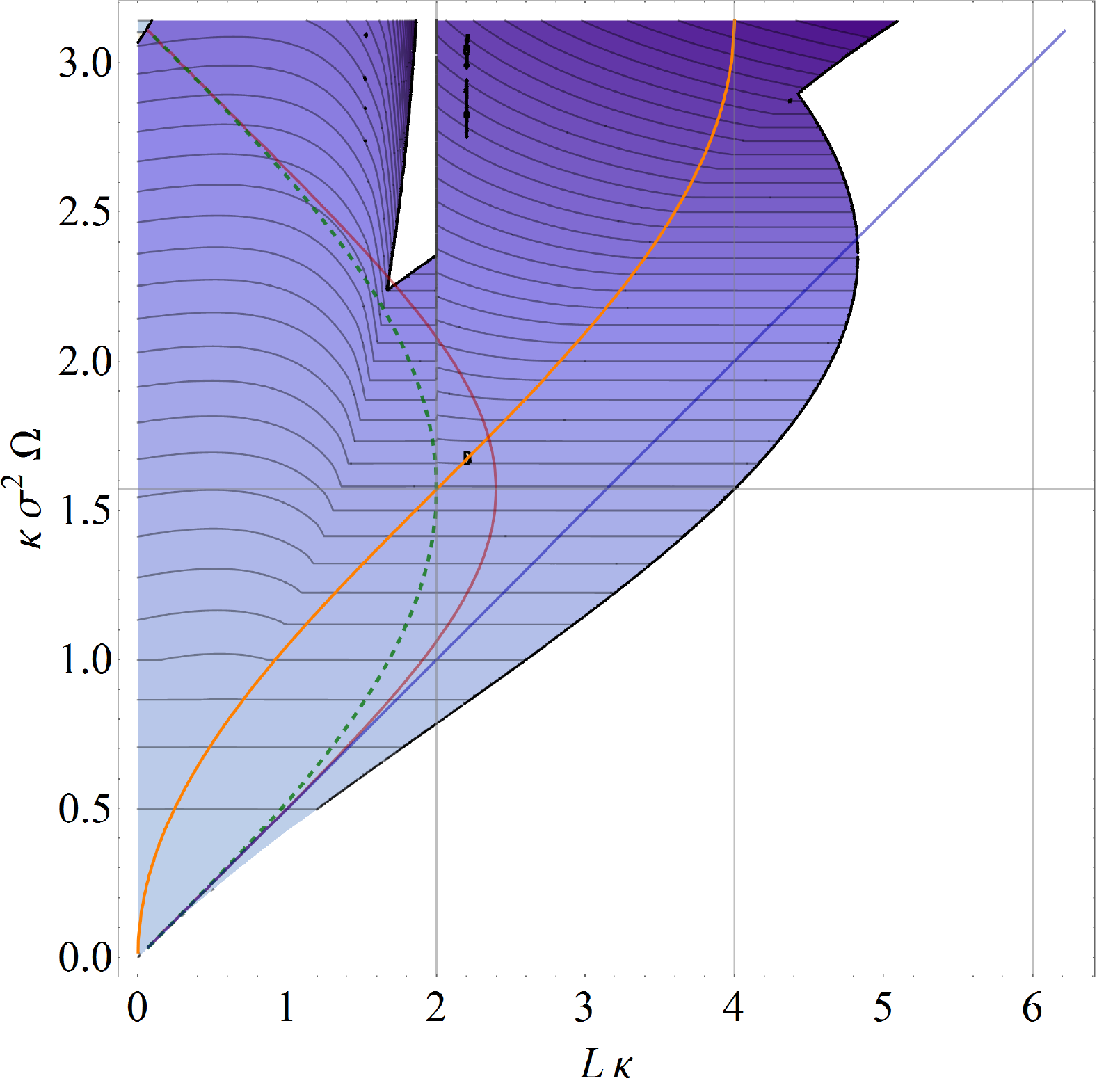}
\par\end{centering}
\caption[Negativity contours for anti-parallel acceleration]{\label{fig:negContours}(Color online.) Regions of nonzero negativity~\cite{Vidal2002} (i.e., parameter regions where entanglement harvesting is possible); see text for meaning of parameters and details of physical setup. \textbf{(Left)} Results from Sec.~\ref{sec:resultspar}. Green region (left of green dotted curve):\ parallel accelerating detectors in Minkowski vacuum ($N_\parsymb > 0$) and also comoving detectors in de~Sitter conformal vacuum ($N_{\text{dS}}>0$). Red+green regions (left of red solid curve):\ inertial detectors in Minkowski thermal bath ($N_{\text{th}}>0$). Green+red+blue regions (above blue solid straight line):\ inertial detectors in Minkowski vacuum ($N_0 > 0$), which can be interpreted as the limit of the other three cases as $\kappa \to 0$. \textbf{(Right)} Negativity profile for anti-parallel accelerated detectors. Harvesting is possible in the blue, contoured region ($N_\antiparsymb >0$). Boundaries of the regions from the left panel are shown overlaid on the right for comparison. The contours within the large blue region are lines of constant $N_\antiparsymb$, with more entanglement toward the bottom. 
The four features discussed in Sec.~\ref{sec:resultsantipar} are illustrated as follows: (1)~The portion of the blue contour region below the solid, blue line is the region of enhancement over inertial detectors (Sec.~\ref{subsec:enhancement}). (2)~The orange line shows the critical distance, Eq.~\eqref{eq:Lcrit}, for entanglement resonance (Sec.~\ref{subsec:resonance}). (3)~What looks like a curvy ``bulge'' on the left-hand side is due to the causal residue contribution (Sec.~\ref{subsec:causalresidue}). (4)~The triangular region in the upper half and with $L\kappa > 2$ corresponds to the noncausal residue contribution (Sec.~\ref{subsec:noncausalresidue}).

}
\end{figure*}

For comparison, the entanglement boundary for inertial detectors in the Minkowski vacuum is given by
\begin{equation}
\label{eq:AnalyticMinkEnt}
	\frac L 2 < \sigma^2\Omega,
\end{equation}
which can be obtained by taking the $\kappa \to 0$ limit of either Eq.~\eqref{eq:AnalyticParaEnt} or Eq.~\eqref{eq:AnalyticThermalEnt} while fixing the other physical parameters $L$, $\sigma$, and~$\Omega$. For entangled states, the negativity in this case is~\cite{Nambu2013}
\begin{align}
\label{eq:NInertial}
	N_0 &\simeq e^{-(\sigma\Omega)^2} \frac{\eta_0^2} {2 \pi} \left[ \left( \frac {\sigma} {L} \right)^2 - \frac {1} {4(\sigma \Omega)^2} \right]\,.
\end{align}
All of these results are shown in Fig.~\ref{fig:negContours}~(left). 
Notice that the negativity is generally exceptionally small since, for all cases considered here, $N \sim e^{-(\sigma\Omega)^2}$.

\section{Results: Anti-parallel acceleration}
\label{sec:resultsantipar}

In order to study anti-parallel acceleration, we choose the negative sign in Eqs.~\eqref{eq:acctrajectories}, giving rise to the trajectories shown on the right of Fig.~\ref{fig:Trajectories}.    Note that the infinite Gaussian tails in the window function~$\eta(\tau)$ cause detectors with overlapping Rindler wedges to be causally connected (not strictly spacelike separated). Nevertheless, as assumed in Ref.~\cite{VerSteeg2009}, we choose the standard deviation~$\sigma \ll L$ so that the nontrivial parts of the interactions are spacelike separated. We calculate $X$ and $A$  using Eqs.~\eqref{eq:saddleX} and~\eqref{eq:saddleA} with $D = D_\antiparsymb$. However, when shifting the contour of integration in $X$, we now do cross a number of poles, and we must calculate the residue contributions separately. As a result, we cannot easily use a simple inequality to represent the entangled region as we did in the previous cases. The details of these residue calculations are given in Appendix~\ref{sec:ResidueCalculation}.

The profile for harvesting entanglement using anti-parallel detectors is shown in Fig.~\ref{fig:negContours}~(right). This plot contains four interesting features, which we describe below, followed by some discussion about which features may be due to the long tails of the Gaussian window function (and would therefore disappear if it were replaced with a similar function having compact support).

\subsection{Enhancement over inertial detectors}
\label{subsec:enhancement}

Perhaps one of the most interesting features of entanglement harvested with anti-parallel acceleration is that the acceleration allows for entanglement harvesting where inertial detectors could not.  In particular, certain detectors with a distance of closest approach $L > 2\sigma^2 \Omega$ can harvest entanglement. This is surprising because it is outside the allowed region for inertial entanglement harvesting, given by Eq.~\eqref{eq:AnalyticMinkEnt}. The allowed region for inertial detectors is represented by the blue triangular region in Fig.~\ref{fig:negContours}~(left), the boundary of which is given by the blue line in Fig.~\ref{fig:negContours}~(right), assuming the same resonant frequency and window function~\cite{VerSteeg2009}. This line should be interpreted as the $\kappa\to 0$ limit of non-inertial motion.

The power of this enhancement can be understood physically.  Consider two inertial detectors in Minkowski spacetime, configured with $L~\gtrsim~2\sigma^2\Omega$---i.e., just below the blue diagonal line in Fig.~\ref{fig:negContours}~(right). These two detectors would not be able to harvest entanglement and would remain in a separable state.  However, if we instead consider two detectors with the same $\Omega$, accelerating oppositely and reaching a distance at closest approach equal to $L$, in some cases they would become entangled.  Notice that we do not have to physically bring the detectors closer together than the original~$L$ in order to entangle them.  It is worth pointing out that this feature arises in the portion of $X$ that does not come from residue terms. Therefore, we conjecture that the enhancement is not an artifact of the infinite tails in the Gaussian window functions but would persist even if the Gaussian were cut off smoothly for $\abs\tau \gg \sigma$ (see Sec.~\ref{subsec:compactwindow}).

For small accelerations and finite interaction times, the two accelerating detectors would look almost like inertial detectors, yet they could in principle harvest entanglement that truly inertial detectors could not. This seems to present a paradox: the behavior as $\kappa \to 0$ should match up with that of actually achieving the limit $\kappa = 0$, but it appears not to do so. This apparent paradox is resolved by noting that for small accelerations, both~$L$ and~$\Omega$ must be very large to see this effect (since the terms $L \kappa$ and $\kappa \sigma^2 \Omega$ both become small), and at the limit point~$\kappa = 0$, $L$ and $\Omega$ must become infinite, which is impossible. Therefore, for all practical purposes, the inertial bound, Eq.~\eqref{eq:AnalyticMinkEnt}, becomes the relevant one in the limit~$\kappa \to 0$.

\subsection{Entanglement resonance}
\label{subsec:resonance}

As mentioned in Sec.~\ref{sec:analysis}, the usual picture employed in the literature for studying accelerated observers~\cite{Unruh1976,Birrell1982,Crispino2008} suggests a natural restriction to trajectories embedded within Rindler wedges sharing a common apex, which corresponds to $L=2\kappa^{-1}$. We do not impose this restriction in this work because there is no physical reason why the separation between two independent objects should necessarily be related to their acceleration.

%
%
This restriction is not entirely without physical motivation, however. The Unruh effect is usually discussed in terms of the Minkowski vacuum being describable as a collection of two-mode squeezed states between pairs of Rindler modes, with one mode from each pair localized to the left and right Rindler wedges, respectively~\cite{Unruh1976,Birrell1982,Crispino2008}.\footnote{Recent results, however, bring the naturalness of this picture into question~\cite{Rovelli:2012jm}.} Due to reflection symmetry between the two wedges, the modes that are entangled (i.e., two-mode squeezed) happen to be those that are naturally detected by anti-parallel-accelerating observers, one within each wedge. For such detectors, $L=2\kappa^{-1}$. 
%
One might therefore wonder if having detectors on these trajectories might somehow be the best way to see the effects of anti-parallel acceleration due to resonance with these two-mode-squeezed modes. In fact, perhaps this configuration might be \emph{necessary} for making use of such a resonant effect to harvest a significant amount of entanglement using anti-parallel-accelerating detectors. Surprisingly, this conjecture will turn out to be false. %
%
%
%

Before we say why this is false, it is worth asking what such a resonant effect would look like in our calculations. We do not expect to see any effect in evaluating~$A$ since this is just the response of our detector to the perceived thermal noise, and this does not depend on the relative positions of the trajectories.  Since the detector is only coupled (nontrivially) to the field for a finite time~$\sigma$, this response is expected to be finite.\footnote{In the case of constant and always-on coupling, $A$ would diverge, which is why it is common to talk about transition rates in that case, rather than the total probability of excitation~\cite{Birrell1982}.} It is still conceivable that we might find a divergence in~$X$, however. Such a divergence would signal a situation in which higher-order terms---representing multiple real or virtual transitions---would be necessary to capture the true dynamics. Since we do not go beyond second order in this analysis, we cannot be sure of what exactly is happening in such a case, but we can still consider such a divergence to be evidence of a resonance condition, which should be detectable in the final state of the detectors. In what follows, we take this interpretation of such divergences.

\begin{figure}[!tb]
\begin{centering}
\includegraphics[width=0.95\columnwidth]{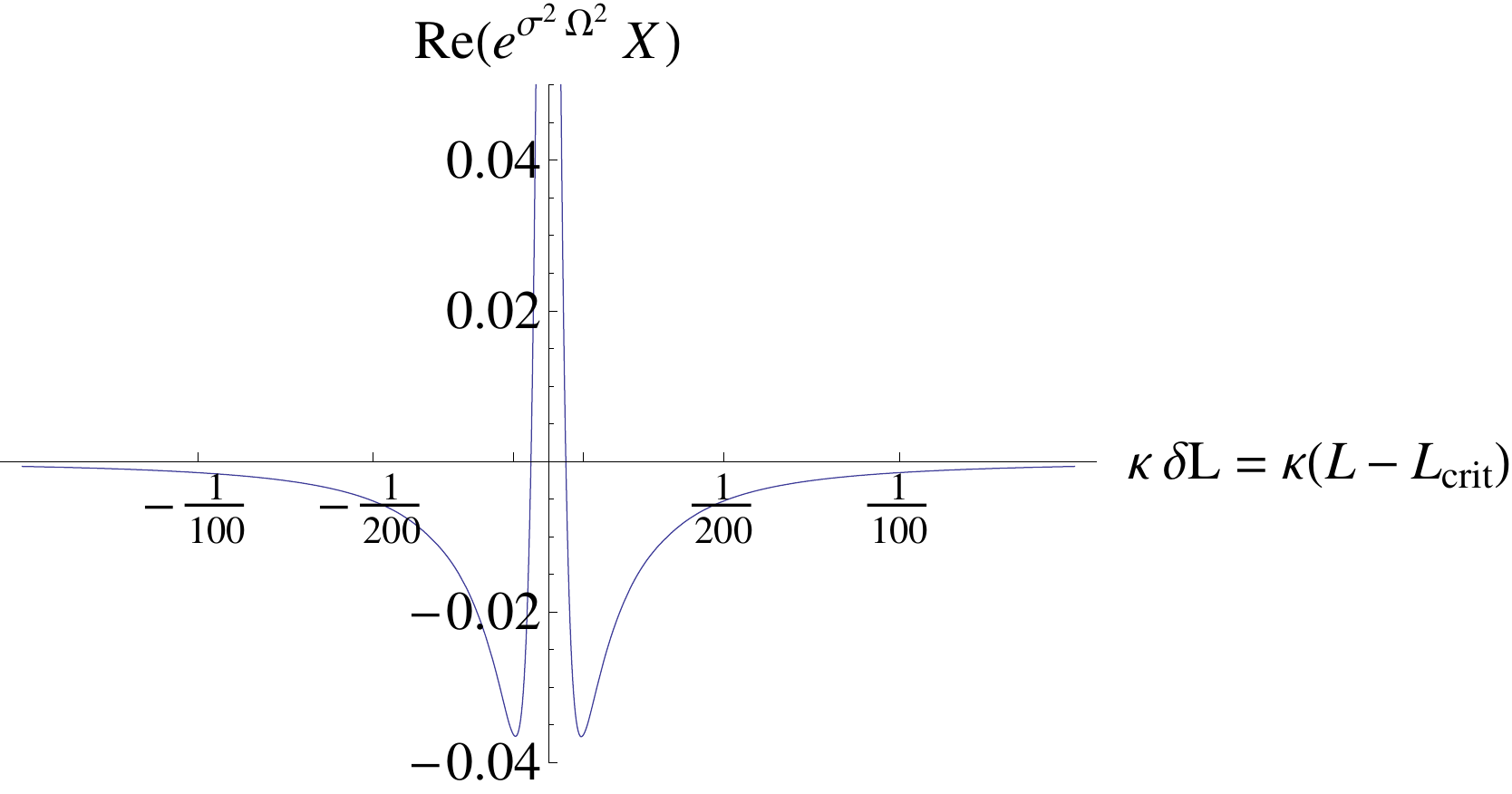}
\par\end{centering}
\caption[$\Re(X)$ as a function of $\delta L$]{\label{fig:formOfX}Behavior of~$\Re(X)$, up to an overall multiplicative factor~$e^{(\sigma\Omega)^2}$, as a function of deviation $\delta L$ from the critical distance~$L_{\text{crit}}$ defined in Eq.~\eqref{eq:Lcrit}, with  ${\kappa=\frac{1}{1000}}$, ${\sigma=1}$, and ${\Omega=1250}$.  The overall shape of the curve is robust to changes in $\Omega$, with only the width changing by a multiplicative factor of order unity.  It can be shown that $\Re(X)<0$ for all values of $\delta L$, except in a small corridor around $\delta L=0$ (i.e.,~centered around $L_\text{crit}$).  Thus, we note that $\Re(X)$ changes sign near the critical distance as the detectors change from correlated to anti-correlated.  This sign change is detectable. See Sec.~\ref{subsec:coherenceCorridor}.}
\end{figure}

It turns out that for any detector resonance frequency~$\Omega$, we can create a situation in which such a resonant condition occurs \emph{for any value of~$L$} up to $4\kappa^{-1}$. To find this \emph{critical distance}~$L_{\text{crit}}$, we evaluate where the saddle-point approximation of~$X$, Eq.~\eqref{eq:saddleX} diverges. Using $D = D_\antiparsymb$, we find
\begin{equation}
\label{eq:Lcrit}
	L_\text{crit}=\frac{2}{\kappa}\left[1-\cos(\kappa\sigma^2\Omega)\right].
\end{equation}
Interestingly, while this method reveals the location of the divergence in~$X$, the saddle-point approximation gets the sign wrong. Direct numerical integration reveals that $X \to \infty$ as $L \to L_\text{crit}$, but the saddle point approximation predicts~$X \to -\infty$ instead. The behavior of $X$ with respect to~$L$ around $L \sim L_\text{crit}$ is shown in Fig.~\ref{fig:formOfX}. Notice that $X$ gets more negative before turning around and shooting up to $+\infty$. The saddle-point approximation does not reveal this turnaround to positive values and instead predicts that $X$ will continue to decrease as the critical distance is approached.

The critical distance is shown as the orange S-shaped curve in Fig.~\ref{fig:negContours}~(right). Given a value of $L < 4\kappa^{-1}$, we can tune the detectors to reveal this effect by setting $\Omega=\Omega_\text{res}$, where
\begin{equation}
\label{eq:Ocrit}
	\Omega_\text{res}=\frac{1}{\kappa\sigma^2}\cos^{-1}\left(1-\frac{L\kappa}{2}\right).
\end{equation}
This divergence appears in the residue-free portion of $X$, which leads us to believe it to be robust to smooth cutoffs in the window functions (see Sec.~\ref{subsec:compactwindow}). An application of this resonance to a rangefinding thought experiment is discussed in Sec.~\ref{subsec:coherenceCorridor}.  Note that we have assumed point-like detectors and that finite size effects could be important for more realistic detectors.


Notice that $L = 2\kappa^{-1}$ is not necessary to see this resonance effect. For the special case~$\kappa\sigma^2\Omega=\frac{\pi}{2}$, of course, we get resonance at that distance, but this is an additional requirement. Therefore, $L = 2\kappa^{-1}$ alone is neither a necessary nor a sufficient condition for resonance. This is contrary to the intuition provided by the Rindler-wedge picture often used to derive and discuss the Unruh effect~\cite{Unruh1976,Birrell1982,Crispino2008}.


%
%
%
%
%
%
%
%
%
%
%
%
%
%
%
%
%
%
\subsection{Causal residue contribution}
\label{subsec:causalresidue}

The entanglement harvested by detectors in different regions of parameter space seems to arise from distinct physical mechanisms.  For $L\kappa<2$, the detectors are timelike separated, and real particles can, in principle, be exchanged. There is a residue contribution to $X$ that is largest for small~$L\kappa$ and vanishes when $L\kappa > 2$. Due to this behavior, we hypothesize that the residue contribution in this region may arise from causal dynamics (see Sec.~\ref{subsec:residueinterp}).

For detectors on the hyperbolic trajectories shown in the right panel of Fig.~\ref{fig:Trajectories}, the Gaussian window functions cause the detectors to spend a very long time traveling near the speed of light and interacting with the underlying field. Although the interaction is very weak, we conjecture that the long interaction time gives rise to a nontrivial contribution to $X$ comparable to (or even greater than) the residue-free portion.

\begin{figure}[!tb]
\begin{centering}
\includegraphics[width=0.8\columnwidth]{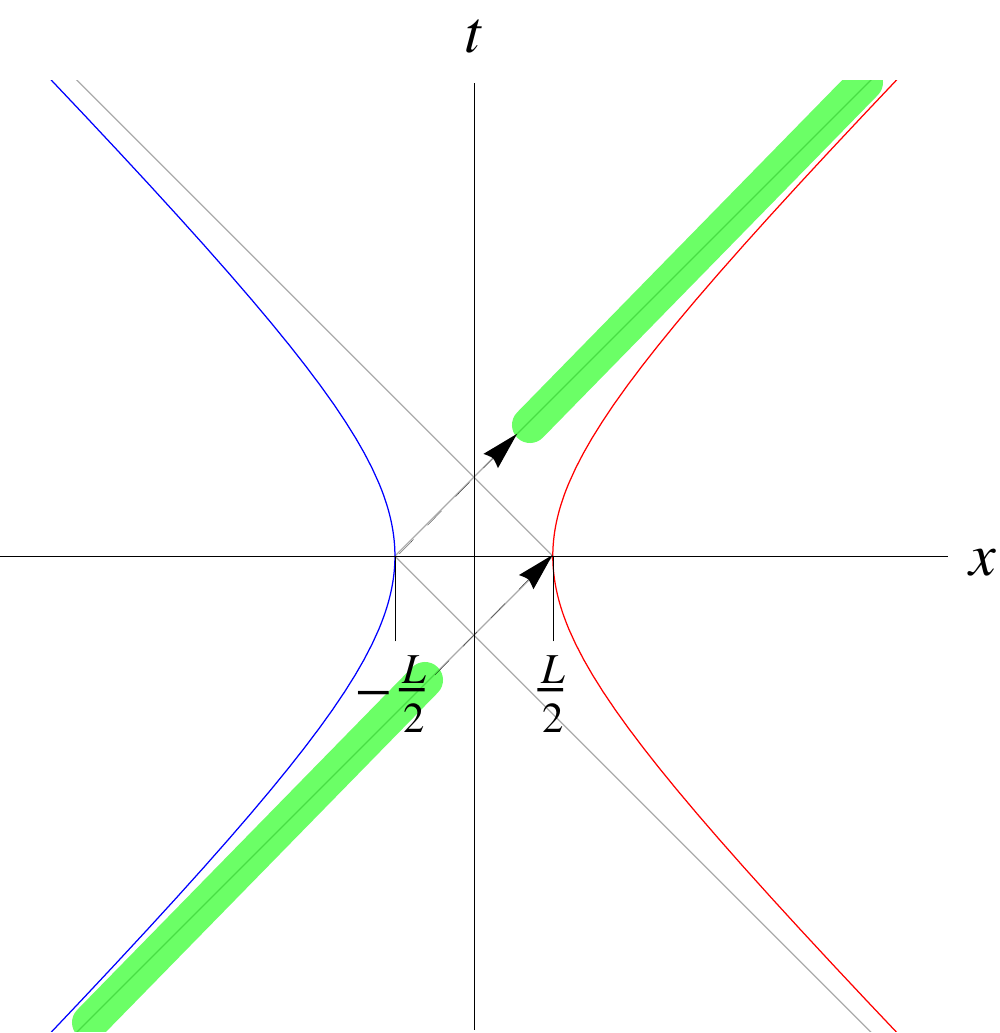}
\par\end{centering}
\caption[Causal residue contribution]{\label{fig:ShockWave}(Color online.)  Interpretation of the residue contribution from causally connected detectors ($L\kappa<2$).  While the analytic form of the Gaussian was essential for calculating the coherence term, $X$, the infinite tails give rise to very long-time interactions with field modes (shown in green) while the detector is traveling near the speed of light.  Effects due to interaction between the first detector and the field over a long period in the infinite past can build up and be felt by the second detector in finite time as it crosses the lightlike extension of the first detector's past asymptote (and vice versa). Similarly, small fluctuations caused by short-time interaction from the first detector can be amplified through long interactions in the infinite future of the second detector (and vice versa). We hypothesize that these effects together give rise to the nontrivial contribution to~$X$ discussed in Sec.~\ref{subsec:causalresidue} and shown on the left side of Fig.~\ref{fig:negContours}~(right).}
\end{figure}

Specifically, we hypothesize that the first detector's interaction with the field for a long time in the infinite past may produce a large effect concentrated near the lightlike past asymptote of the first detector's trajectory that is felt by the other detector (in the coherence term~$X$) as the second detector crosses the future extension of that asymptote. (The same effect also happens from the second to the first detector.) Similarly, the second detector will spend a long time near the lightlike future asymptote of its own trajectory, thus amplifying any effect due to the presence of the first detector near the past extension of that asymptote (and vice versa).
The net result of these four (symmetric) processes is a nontrivial contribution to the coherence term~$X$, which indicates that the detectors ``feel each other's presence'' due to the long tails of the Gaussian and overlapping wedges,
as shown in Fig.~\ref{fig:ShockWave}.

\subsection{Noncausal residue contribution}
\label{subsec:noncausalresidue}

Even as the causal residue contribution discussed in Sec.~\ref{subsec:causalresidue} vanishes as $L\kappa > 2$, a second residue contribution takes over in that region when, in addition, $\kappa\sigma^2 \Omega > \tfrac \pi 2$. This means that there exists a nontrivial residue contribution for spacelike separated detectors ($L\kappa>2$) but only for certain choices of the other parameters. Because the detectors are spacelike separated for the entirety of their trajectories in this case, this contribution is more mysterious. There are some clues that we can use to hypothesize about its physical origin, however.

The most important clue is that the effect is strongest near $L\kappa \gtrsim 2$. This suggests that the detectors may be feeling the effects of entangled Rindler modes~\cite{Unruh1976,Birrell1982,Crispino2008}. To understand this reasoning, consider dividing spacetime into four wedges (using the usual Rindler prescription~\cite{Unruh1976}) whose common origin is at ${(x=0, t=0)}$. If the actual detector trajectories have asymptotes that are close to the boundaries of these origin-centered Rindler wedges (i.e.,~${L\kappa \gtrsim 2}$), then the detectors will be nearly resonant with the Rindler modes that are two-mode squeezed (see Sec.~\ref{subsec:resonance}). This effect only increases as $L\kappa \to 2^+$. However, the fact that it cuts off sharply as soon as $L\kappa < 2$ also suggests that the infinite tails may play a role in this effect. Perhaps confinement of the infinite tails to the left and right origin-centered wedges is required to see the effect. If the detectors are too close, then the long tails of each detector's trajectory will leak out into the origin-centered future and past wedges. This effect would then be traded out for the causal effect described in Sec.~\ref{subsec:causalresidue}, which is tiny for $L\kappa \lesssim 2$. Further work is needed to explore the effect of using switching functions with compact support, but this requires new calculational methods and is beyond the scope of this work. We therefore leave the above as a working hypothesis. Further justification may be found in Sec.~\ref{subsec:residueinterp}.

\section{Rangefinding}
\label{sec:range}

Another interesting feature of the harvesting process is that it depends critically on the distance between the two detectors.  Entanglement harvesting depends on the interplay between $\Omega$, $L$, and $\kappa$, and we will present three thought experiments for \emph{rangefinding} (i.e., measuring the distance of closest approach between two anti-parallel accelerating detectors) using the properties of the harvested entanglement and their dependence on the relevant parameters.  The goal is not to propose viable methods for measuring distance but rather to expose the relationship between distance and harvested entanglement in the hopes of motivating future studies into practical applications of this phenomenon.

The first of the three techniques (described in Sec.~\ref{subsec:coherenceCorridor}) makes use of the critical distance described in Sec.~\ref{subsec:resonance} and uses the divergence in $X$ to determine the distance between the detectors.  The second and third techniques rely on sudden death and revival of entanglement.  Here, ``death'' and ``revival'' of entanglement are to be understood through ensembles of detector configurations constructed such that they form a path in parameter space along which the entanglement between such detectors exhibits extremely rapid changes.  We are not referring to death and revival of entanglement with respect to time evolution.  This distinction is emphasized in Sec.~\ref{subsec:deathrevival}. We will describe a method for using sudden death of entanglement as a signal for crossing a specific distance, and a method of using steep gradients in the negativity (revival of entanglement) to give precise information about \emph{changes} in distance. 

\subsection{Coherence Corridor}
\label{subsec:coherenceCorridor}

Having found the critical distance in Eq.~\eqref{eq:Lcrit}, we explore the properties of the harvested entanglement near $L_\text{crit}$.  We will study the dependence of $\Re(X)$ on $L$.  For convenience, we define 
\begin{equation}
\delta L=L-L_\text{crit}
\end{equation}
to be the distance away from the critical value.  Fig.~\ref{fig:formOfX} shows $\Re(X)$ as a function of $\delta L$ for $\kappa=0.001$, $\sigma=1$, and $\Omega=1250$ (i.e., $\kappa\sigma^2\Omega=1.25$).  These parameter values were chosen to lie in a region of parameter space in which residue contributions to $X$ are negligible; specifically, we choose $1.2<\kappa\sigma^2\Omega<\tfrac \pi 2$ and $1.1<L\kappa<2$.  In this region of parameter space $X$ is heavily dominated by the residue-free contribution (i.e., the residue terms are negligible), and the shape of $\Re(X)$ is generic for other values of $\Omega$.

Through explicit numerical evaluation we find that the coherence term, $X$, flips sign in a narrow corridor around $L=L_\text{crit}$. (Interestingly, the saddle-point approximation of Eq.~\eqref{eq:saddleXantipara} misses this behavior.)  This sign flip is detectable by performing local measurements on the detectors and collecting statistics.  To see how this is possible, consider the bipartite state of the two-detector system in the basis defined by the tensor product of their respective ground and first excited states, up to second order in the detector coupling.  The density matrix is of the form
\begin{equation}
\rho_{ab}= 
\begin{pmatrix}
C & 0 & 0 & -X^{*}\\
0 & A & B^* & 0\\
0 & B & A & 0\\	
-X & 0 & 0 & 1-2A-C
\end{pmatrix}
\,.
\end{equation}
By measuring $\sigma_x\ox\sigma_x$ or $\sigma_y\ox\sigma_y$, we have that
\begin{align}
\label{eq:}
	\langle\sigma_{x}\otimes\sigma_{x}\rangle &= -2\Re(X)+2\Re(B),\\
	\langle\sigma_{y}\otimes\sigma_{y}\rangle &= 2\Re(X)+2\Re(B).
\end{align}
Therefore, 
\begin{equation}
\label{eq:ReX}
4\Re(X)=\langle\sigma_{y}\otimes\sigma_{y}\rangle-\langle\sigma_{x}\otimes\sigma_{x}\rangle.
\end{equation}

Operationally, this means that we can make measurements of either ${\sigma_x\otimes\sigma_x}$ or ${\sigma_y\otimes\sigma_y}$ and send the results of the measurements, along with a timestamp and information about the choice of measurement, back to a neutral third party at $x=0$.  If the third party finds that the measurements were made in the same basis, then the results of the measurements are kept.  With access to an ensemble of detectors and measurements, one can collect statistics of the outcomes of the measurements for comparison.  Using this scheme, it is possible to use entanglement to measure distance between the detectors.   It is in this sense that we say information about the distance between the detectors is stored nonlocally in the phase of the joint state.

Suppose for a moment that we had access to such an ensemble of detectors with $1.2<\kappa\sigma^2\Omega<\tfrac \pi 2$, and we performed the measurements described above.  In this case, $\Re(X)$ can only be positive when $L$ is close to $L_\text{crit}$ (i.e., $\delta L \simeq 0$).  As such, a measurement outcome of $\Re(X)>0$ necessarily implies that the detectors must have been separated by $L_\text{crit}$ up to $\mathcal{O}(\delta L)$.

\subsection{Sudden Death of Entanglement and a Relation to Distance}
\label{subsec:deathrevival}

The negativity contours [Fig.~\ref{fig:negContours}(Right)] for anti-parallel acceleration show a narrow region for large $\kappa\sigma^2\Omega$ where the negativity is zero.  We have affectionately dubbed this oddly shaped region the ``necktie'', for obvious reasons.  Just to the right of the necktie (e.g., $L\kappa \gtrsim 2$ and $\kappa\sigma^2\Omega\gtrsim 2.4$) the negativity suddenly drops to zero as we cross $L\kappa=2$ from the right.  This sudden death of entanglement (in parameter space) is curious in its own right and has been discovered in a number of other systems\citep{Yu2009,Ostapchuk:2011ud,Doukas:2013noa}. In addition, we can use this phenomenon for rangefinding by identifying the presence or absense of entanglement as a signal for crossing a particular distance.

To see how entanglement sudden death can be used for rangefinding, we shall construct a simple toy protocol that uses entanglement as a trigger for signalling that a particular distance ($L=2\kappa^{-1}$) has been crossed.  For the remainder of this section, we will always require $\kappa\sigma^2\Omega\gtrsim~2.4$ in order to focus on the necktie.

Consider two sets of detectors separated by $L={2 \kappa^{-1}\pm\delta}$ for $0<\delta\ll 1$.  The detectors separated by $L={2 \kappa^{-1}+\delta}$ will be able to harvest entanglement and will evolve into an entangled state, whereas the detectors separated by $L=2 \kappa^{-1}-\delta$ will not be able to harvest any entanglement and will remain in a separable state.  
Using an ensemble of detectors with varying $L$, it is possible to use entanglement to identify those detectors lying on either side of the $L\kappa=2$ boundary.  This identification marks the desired $L=2\kappa^{-1}$ distance.  In order to verify that the detectors have evolved into an entangled state, it suffices to show violation of a Bell inequality, which can always be done if and only if the state is entangled (with sufficient extra resources) \cite{Masanes:2008ib}.

\subsection{Negativity Gradient}
\label{subsec:neggrad}

In addition to the sudden death of entanglement, there is another feature of the necktie region that we can use for rangefinding.  To reiterate an earlier point, our aim here is simply to highlight the relationship between entanglement and distance.  From Fig.~\ref{fig:negContours}(Right), we see that the negativity has a steep gradient in the region just to the left of the necktie.  In this region, the negativity changes rapidly over very small changes in distance~$L$.  As such, the amount of entanglement harvested in this region of parameter space is highly sensitive to the distance between the detectors; very small changes in the separation distance $L$ manifest as (proportionally) large changes in the negativity.  We can use this feature to determine \emph{changes} in distance very precisely by measuring relatively coarse changes in the entanglement.

To make this argument precise, suppose we have two pairs of detectors with a given acceleration $\kappa$ and energy gap $\Omega$.  Suppose further that the pairs are both separated by a known distance at closest approach $L$.  If the separation between one pair of detectors is perturbed slightly, the negativity for that pair will be vastly different from the negativity of the unperturbed, reference pair.  If we measure~$A$ and~$X$ for the two pairs of detectors [the latter through coincidence measurements as in Eq.~\eqref{eq:ReX}] and compare them, we can map the change in negativity to a change in distance.  The steep negativity gradient allows even relatively imprecise knowledge of the negativities to correspond to fairly precise values of distance.  Note that this technique does not allow us to measure arbitrary absolute distances, but rather fluctuations in distance around some target value.

\subsection{Caveat}

\label{subsec:rangefindingcaveat}

Obviously, much more practical methods exist to measure distance between objects. Furthermore, the amount of entanglement harvested using these methods is extremely small due to the Gaussian prefactor~$e^{-(\sigma\Omega)^2}$, so even the proportionally large changes in entanglement predicted in, e.g., Sec.~\ref{subsec:neggrad} are between values that are exceedingly small.

As such, we would like to remind the reader that the purpose of this section is simply to outline several thought experiments to show---\emph{in principle}---how the particular correlations in the harvested entanglement reveal details about the distance between the traveling detectors. We expect that this effect---of the harvested entanglement showing a critical dependence on distance---is generic and will eventually lead to realistic proposals for rangefinding using entanglement harvesting. One example of such a proposal is Ref.~\cite{Brown:2014en}, which was motivated directly by the results reported here. 

\section{Summary}
\label{sec:summary}

Entanglement harvesting is the process of swapping entanglement out of a quantum field and into two Unruh-deWitt detectors using local interactions with the field.  Here we studied the entanglement harvested by uniformly accelerating detectors on various trajectories.  For detectors accelerating in opposite directions, we broke away from the usual approach applied to Rindler observers, wherein the observers have commensurate Rindler wedges meeting at the origin and have a distance at closest approach that is set by their acceleration parameter ($L=2\kappa^{-1}$).  We instead allowed the detectors an arbitrary minimum separation---a more physically meaningful approach to the problem.  While the usual approach is well motivated by the fact that the Minkowski vacuum corresponds to a two-mode squeezed state of Rindler modes (which are the natural modes detected by observers with $L=2\kappa^{-1}$), resonant interaction with these modes is neither necessary nor sufficient for entanglement harvesting.  Instead, we found a relationship between the distance $L$ and energy gap $\Omega$ such that the detectors display a resonant spike in the harvested entanglement, just as they would at $L=2\kappa^{-1}$ and $\kappa\sigma^2\Omega=\pi/2$.  This resonance defines a critical distance at which the Wightman function in Eq.~\ref{eq:saddleX} diverges.

An additional benefit of promoting $L$ to a free parameter is the emergence of an enhancement over inertial motion in entanglement harvesting.  A variable $L$ opens up a previously inaccessible region of parameter space in which entanglement harvesting is possible.  This enahancement is a feature of the nonrelativistic motion and persists for small $\kappa$.  However, in order to observe this effect with small accelerations we would need very large distances $L\sim\mathcal{O}(4\kappa^{-1})$.

In order to compute the coherence term $X$, we exploited properties of the window functions governing the detector-field interaction.  We chose analytic window functions (Gaussian, to be precise) so that we could split our calculation into a residue-free portion and residue terms.  The residue-free contribution is robust to changes in the window functions, while the residue contributions are conjectured to be related to the infinite Gaussian tails.  We found two major residue contributions to the coherence term: one arising from apparently causal dynamics and another from apparently noncausal dynamics.  The noncausal residue contribution arises whenever the detectors lie in completely disjoint Rindler wedges so that no null or time-like communication could ever occur.  In this case, we found a large residue contribution, and we hypothesized about its origin.  We also described the causal residue contribution for which the Rindler wedges of the two detectors have some overlap and communication could occur.  We leave for future work an analysis of these contributions with compact window functions (see Sec.~\ref{subsec:compactwindow}).

In order to emphasize the relationship between harvested entanglement and distance, we presented three techniques for rangefinding using the aforementioned properties of harvested entanglement from detectors accelerating in opposite directions at the same rate.  The first rangefinding technique relies on the unique shape of the coherence term (i.e, the resonance) near the critical distance.  In a narrow range of $L$ near the critical distance the residue-free portion of $\Re(X)$ undergoes an observable sign change from negative to positive.  We presented a simple LOCC protocol for detecting this sign flip and using the detection as a method of measuring distance.  A second rangefinding technique makes use of the sudden death of entanglement for $\kappa\sigma^2\Omega\gtrsim~2.4$ by using the death of entanglement to trigger on a particular distance.  Since the negativity  drops suddenly to zero at a specific distance ($L=2\kappa^{-1}$), the presence or absence of entanglement on either side of this boundary identifies the transition distance.  Finally, the third rangefinding technique we presented serves as a means for sensitively detecting fluctuations in distance around a known reference value.  Using a sharp gradient in negativity  over short distances, large deviations in negativity away from a known value map to very small deviations in distance away from the reference.  With a reliable method of measuring entanglement, this technique could be used as a seismometer or for maintaining precise distance control. This possibility has motivated a recent proposal in such a direction~\cite{Brown:2014en}. 

In addition to anti-parallel acceleration, we studied entanglement harvesting from detectors accelerating in the same direction at the same rate.  We found a familiar degradation of entanglement for these detectors, and we showed that the degradation is identical to that between two comoving detectors in de Sitter space, yet different from that of inertial detectors in a thermal bath!\cite{VerSteeg2009}.  For parallel acceleration, we found no residue contributions to the coherence term.  As such, we expect our results in this case to hold if we change the window functions to have compact support.

Entanglement harvesting is highly dependent on field and detector parameters~\cite{Reznik2003,Reznik2005,VerSteeg2009,Nambu2011,Nambu2013}. Here we have demonstrated that antiparallel acceleration produces surprisingly rich results not seen in other correspondingly simple cases. Still, many questions remain. For instance, previous work has shown that a minimally coupled field produces a very different result than conformal coupling in an expanding scenario~\cite{Nambu2011,Nambu2013}. Interesting extensions therefore include entanglement harvesting with massive fields, higher-spin fields, and other states of the field---e.g., the Bunch-Davies vacuum~\cite{Bunch1978}, which has important applications in cosmology. We hope the work presented here inspires further studies in this direction.
\blk

\acknowledgments

{We thank Eduardo Mart\'in-Mart\'inez, Achim Kempf, and Eric Brown for discussions. This work was supported in part by the Natural Sciences and Engineering Research Council of Canada and by the Australian Research Council under grant No. DE120102204.  Research at Perimeter Institute is supported by the Government of Canada through Industry Canada and by the Province of Ontario through the Ministry of Research \& Innovation. 
}

\appendix

\section{Complex Integration}\label{sec:ResidueCalculation}

\begin{figure}[!t]
\begin{centering}
\includegraphics[width=0.8\columnwidth]{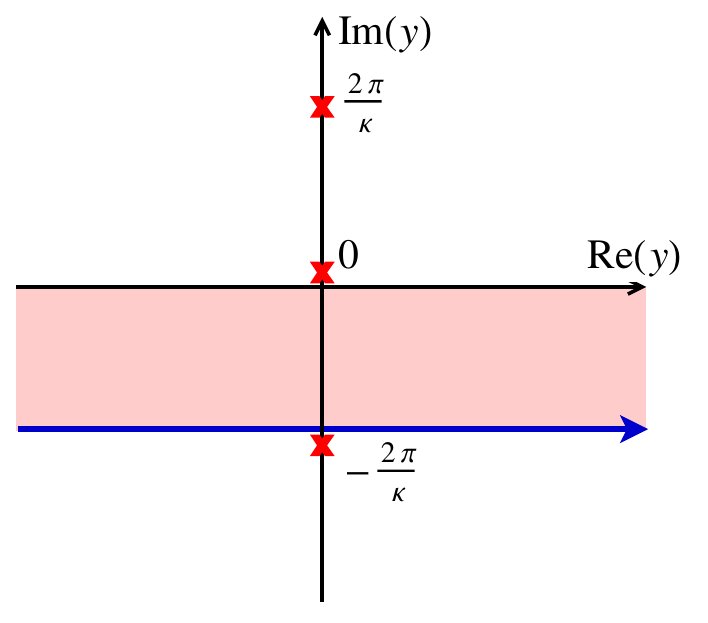}
\par\end{centering}
\caption[Pole structure of D_\text{Detect}]{\label{fig:APoles}The pole structure of $D_\text{Detect}$ showing the shifted contour of integration.  If we restrict $\kappa\sigma^2\Omega<\pi$, we never cross any poles in the evaluation of $e^{(\sigma\Omega)^2}A$, and there are no additional residue terms.  The resulting integral can be computed using the method of steepest descent.  Note that the 
pole locations shown in the plot should be interpreted as being raised slightly by $+i\epsilon$.}
\end{figure}

In order to evaluate the integrals in Eqs.~\eqref{eq:A} and~\eqref{eq:Xeq}, we shift the contour of integration and use Cauchy's residue theorem.  Focusing on $A$ for a moment, we are interested in an integral of the form 
\begin{equation}
A=\frac{\eta_0^2}{2}\int_{-\infty}^{\infty}e^{-\frac{x^2}{4\sigma^2}}dx\int_{-\infty}^{\infty} e^{-\frac{y^2}{4\sigma^2}}e^{-i\Omega y}D_{\text{detect}}(y) dy,
\end{equation}
for overall coupling constant $\eta_0 \ll 1$.  Since we are comparing this quantity with an analogous quantity ($X$), we are free to eliminate a common factor of $e^{-(\sigma\Omega)^2}$ without affecting the entanglement criterion.  Multiplying through and performing the $x$ integral leaves 
\begin{equation}
	e^{(\sigma\Omega)^2}A=\eta_0^2\sqrt{\pi}\sigma\int_{-\infty}^{\infty} e^{-\frac{y^2}{4\sigma^2}}e^{-i\Omega y+\sigma^2\Omega^2} D_{\text{detect}}(y) dy\,.
\end{equation}
This completes the square in the exponential, giving
\begin{equation}
	e^{(\sigma\Omega)^2}A=\eta_0^2\sqrt{\pi}\sigma\int_{-\infty}^{\infty} e^{-\frac{(y+2i\sigma^2\Omega)^2}{4\sigma^2}} D_{\text{detect}}(y) dy\,,
\end{equation}
which is a Gaussian integral with a \emph{complex} mean.  To remove this complex mean, we shift the contour down in the complex plane by an amount $2i\sigma^2\Omega$.  We then define a new variable $y' = y+2i\sigma^2\Omega$ and subsequently drop the prime.  The change of variables removes the complex mean from the exponential and gives us a final integral of the form
\begin{align}
\label{eq:Anoexpres}
	e^{(\sigma\Omega)^2}A&=\eta_0^2\sqrt{\pi}\sigma\int_{-\infty}^{\infty} e^{-\frac{y^2}{4\sigma^2}} D_{\text{detect}}(y-2i\sigma^2\Omega) dy \nonumber \\
	&\qquad-2\pi i \sum_{\text{poles}}(\text{residues})
\end{align}
Since we shifted the contour of integration, residue contributions potentially appear, which are noted explicitly in Eq.~\eqref{eq:Anoexpres}. Since we have shifted it downward in the complex plane, the new contour would include tiny \emph{clockwise} circles around any poles, which accounts for the $-$~sign in front of the residue contributions.

The pole structure of the integrand is shown in Fig.~\ref{fig:APoles}.  Since the exponential is entire, the poles come from $D_\text{detect}$, and we find that they occur at $y_\text{pole}=-\tfrac{2\pi i}{\kappa}n + i\epsilon$ for $n\in\mathbb{Z}$.  Since we only shift the contour by $2i\sigma^2\Omega$, we never cross any poles if we limit $\kappa\sigma^2\Omega<\pi$, and thus the residue contribution vanishes. We therefore restrict our analysis to this case for simplicity. We then evaluate the integral using the method of steepest descent, as in Eq.~\eqref{eq:saddleA}.  For parallel acceleration, a similar method works for evaluating $e^{(\sigma\Omega)^2}X$ using Eq.~\eqref{eq:saddleX}.

In contrast to parallel acceleration, anti-parallel acceleration \emph{does} give rise to poles that are crossed when shifting the contour of integration.  The result for $A$ does not change from the above, but extra care must be given to the calculation of $X$.  

We begin by multiplying $X$ by $e^{(\sigma\Omega)^2}$, which completes the square in the exponent of the integrand:
\begin{equation}
	e^{(\sigma\Omega)^2}X=-\eta_0^2\int_{0}^{\infty}dy\int_{-\infty}^{\infty}dx\, e^{-\frac{y^2+(x-2i\sigma^2\Omega)^2}{4\sigma^2}}D_\antiparsymb(x,y)\,.
\end{equation}
We now shift the contour of integration \emph{up} in the complex plane by $2i\sigma^2\Omega$,  define $x' = x-2i\sigma^2\Omega$, and drop the prime.  The full integral is then
\begin{align}
\label{eq:Xwithresidues}
	e^{(\sigma\Omega)^2}X&=-\eta_0^2\int_{0}^{\infty}\!dy\int_{-\infty}^{\infty}\!dx\thinspace e^{-\frac{x^2+y^2}{4\sigma^2}}D_\antiparsymb(x+2i\sigma^2\Omega,y) \nonumber \\
	&\qquad +2\pi i \sum_{\text{poles}}(\text{residue integrals})\,,
\end{align}
where the residue terms are integrals over $y$.  We evaluate the residue-free portion of the integral using the method of steepest descent, as with the parallel case.  The residues require some work, and their evaluation is described 
next.

\subsection{Residue integral evaluation (anti-parallel case)}

We must consider residue contributions whenever $D_\antiparsymb (x,y)$ diverges for complex~$x$, which occurs iff
\begin{align}
\label{eq:Xpolecond}
	e^{\pm\frac{y\kappa}{2}} \left(\frac {L\kappa} {2}-1\right) \pm i\epsilon + \cosh \left(\frac{x \kappa}{2}\right) = 0\,.
\end{align}
This is actually two conditions, one corresponding to the top sign~($+$) and one to the bottom sign~($-$). Only one condition can be satisfied at a time, so the poles are always simple. Defining $b\coloneqq1-\frac{L\kappa}{2}$, we can solve this equation formally for $x$ to find the pole locations as a function of~$y$ (since we must integrate over~$y$ later):
\begin{align}
\label{eq:xpole}
	x^\pm_{\text{pole}}(y) = \frac 2 \kappa \arccosh \bigl(b e^{\pm\frac{y\kappa}{2}} \mp i\epsilon \bigr)\,.
\end{align}
Note that this function is multivalued by virtue of the multivalued nature of the complex $\arccosh$. 
The relevant pole locations are those for which $0 < \Im x_{\text{pole}}^\pm < 2\sigma^2 \Omega$ (recall that the integration contour for $x$ was shifted upward).

\begin{figure}[!t]
\begin{centering}
\includegraphics[width=0.8\columnwidth]{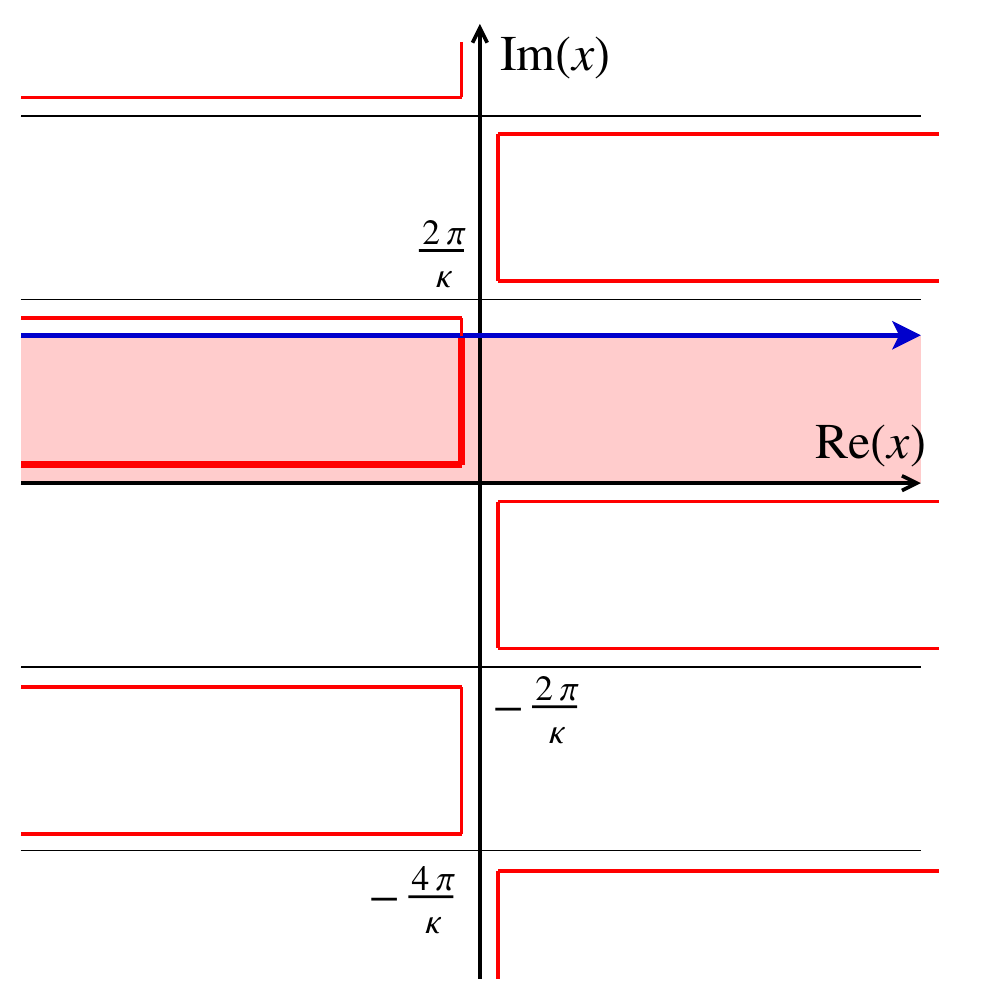}
\par\end{centering}
\caption[Pole structure of D_\antiparsymb]{\label{fig:XPoles}Illustration of the pole structure of $D_\antiparsymb$, along with the up-shifted contour of integration (blue arrow). The collection of horseshoe-shaped red curves is the solution to Eq.~\eqref{eq:Xpoleimag} using the $+$~sign. (Solution curves for the $-$~sign are just the left-right mirror image of these.) Within this solution set, dependence of the pole locations on $L$ (through $b$) and on $y$ is obtained by also solving Eq.~\eqref{eq:Xpolereal}. 
Since we integrate over $y$ in finding~$X$, the location of the poles moves along the red lines shown above. Restricting $\kappa\sigma^2\Omega<\pi$ ensures that we only have to worry about poles on the L-shaped thick red line between the up-shifted contour and the real-$x$ axis.  Note that the $\epsilon$ prescription is responsible for shifting the poles infinitesimally off the real and imaginary axes as shown.}
\end{figure}

To understand the structure of the pole locations, we will take the real and imaginary parts of Eq.~\eqref{eq:Xpolecond} separately. Writing $x = x_\rpart + i x_\ipart$, we obtain the following necessary and sufficient condition for a pole:
\begin{align}
\label{eq:Xpolerealimag}
	b e^{\pm\frac{y\kappa}{2}} = \cosh \left[\frac{(x_\rpart + i x_\ipart) \kappa}{2}\right] \pm i\epsilon\,.
\end{align}
Taking the imaginary part eliminates~$b$ and~$y$ (since $y$ will be integrated along the real axis), giving
\begin{align}
\label{eq:Xpoleimag}
	0 &= \sin \left(\frac{x_\ipart \kappa}{2}\right) \sinh \left(\frac{x_\rpart \kappa}{2}\right) \pm \epsilon \,.
\end{align}
The solutions to this condition for the $+$ sign are plotted in Fig.~\ref{fig:XPoles}; the solutions for the $-$ sign are just the left-right mirror image of this. Poles must also satisfy the condition resulting from the real part, namely
\begin{align}
\label{eq:Xpolereal}
	b e^{\pm\frac{y\kappa}{2}} &= \cos \left(\frac{x_\ipart \kappa}{2}\right) \cosh \left(\frac{x_\rpart \kappa}{2}\right)\,.
\end{align}

The only poles we care about require, in addition, that $x_\ipart < 2 \sigma^2 \Omega$, which translates to $\cos(\frac {x_\ipart \kappa} {2}) > \cos(\kappa \sigma^2 \Omega)$ since we are only considering the case where $\kappa \sigma^2 \Omega < \pi$. Examining Eq.~\eqref{eq:Xpoleimag} or Fig.~\ref{fig:XPoles}, we see that the poles in question always ``hug'' either the real or imaginary axis. This means that if $x_\ipart > 0$ then $x_\rpart = 0^\mp$. We can use these facts and Eq.~\eqref{eq:Xpolereal} to write the following necessary and sufficient condition for inclusion of a pole in the residue contribution:
\begin{align}
\label{eq:Xpoleinclusion}
	be^{\pm\frac{y\kappa}{2}} > \cos(\kappa\sigma^2\Omega)\,.
\end{align}

By lifting the contour upward, we deform it to include tiny \emph{counterclockwise} circles around the poles. Therefore, the contribution labeled ``residue integrals'' in Eq.~\eqref{eq:Xwithresidues}, which consists of residues integrated over $y$, is added with a $+$~sign. In particular,
\begin{align}
\label{eq:rescontrib1}
	2\pi i & \sum_{\text{poles}} \binom {\text{residue}} {\text{integrals}} \nonumber \\ 
	&= 2 \pi  i \sum _{s \in \{+,-\}} \int_0^{\infty } dy\, \Theta \Bigl[be^{s \frac{y\kappa}{2}} - \cos(\kappa\sigma^2\Omega) \Bigr] R_s(y) \,,
\end{align}
where $\Theta(x)$ is the Heaviside step function, which is used to enforce the pole inclusion condition, Eq.~\eqref{eq:Xpoleinclusion}, and $R_s(y)$ is the residue of the integrand in Eq.~\eqref{eq:Xwithresidues}. Explicitly,
\begin{align}
\label{eq:residue}
	R_\pm(y) &\coloneqq \res_{x = x^\pm_{\text{pole}}(y)} \left[ -\eta_0^2 e^{-\frac{y^2+(x-2i\sigma^2\Omega)^2}{4\sigma^2}}D_\antiparsymb(x,y) \right] \nonumber \\
	&= -\eta_0^2 e^{-\frac{y^2+(x^\pm_{\text{pole}}-2i\sigma^2\Omega)^2}{4\sigma^2}} \res_{x = x^\pm_{\text{pole}}(y)} D_\antiparsymb(x,y)\,,
\end{align}
where we now force the $\arccosh$ in Eq.~\eqref{eq:xpole} to produce a unique answer by restricting its range to have imaginary part $\in (0, \pi)$ (with the endpoints of this interval excluded since ${\epsilon > 0}$), and the second line follows because the exponential is an entire function and the poles are all simple. It will be useful to define two new variables in terms of~$y$:
\begin{align}
\label{eq:thetapm}
	\theta_\pm(y) \coloneqq \arccosh \bigl(b e^{\pm\frac{y\kappa}{2}} \mp i\epsilon \bigr)\,, \quad \Im \theta_\pm \in (0,\pi)\,,
\end{align}
where the restriction on the range of $\arccosh$ ensures that these functions are single valued. In these variables,
\begin{align}
	D_\antiparsymb &= \frac{\kappa^2}{16\pi^2}
	\left[\cosh \left(\frac{x \kappa}{2}\right) - \cosh \theta_- \right]^{-1}
	\nonumber \\ &\quad\quad \times
	\left[\cosh \left(\frac{x \kappa}{2}\right) - \cosh \theta_+ \right]^{-1}.
\end{align}
The pole condition, Eq.~\eqref{eq:Xpolecond}, becomes simply
\begin{align}
\label{eq:Xpolecondtheta}
	x^\pm_{\text{pole}}(y) = \frac 2 \kappa \theta_\pm(y)\,.
\end{align}
Using this and the usual formula for the residue of a simple pole, we can evaluate
\begin{align}
\label{eq:residueD}
	\res_{x = x^\pm_{\text{pole}}(y)} D_\antiparsymb &= \frac {\kappa} {8\pi^2} \frac {\csch \theta_\pm} {\cosh \theta_\mp - \cosh \theta_\pm}
	\,.
\end{align}
Having served its purpose, we now let $\epsilon \to 0^+$ for the rest of the evaluation.

To perform the integral over $y$, we change variables to $\theta_\pm$ as appropriate, along with
\begin{align}
	dy &= \pm \frac 2 \kappa \tanh \theta_\pm\, d\theta_\pm \,, \\
	y &= \pm \frac 2 \kappa \log \left(\frac {\cosh \theta_\pm} {b} \right)\,.
\end{align}
This allows us to rewrite the $y$-integral in Eq.~\eqref{eq:rescontrib1} as 
\begin{widetext}
\begin{align}
\label{eq:rescontribintegral}
	&\int_{\theta_s(0)}^{\theta_s(\infty) } s \frac 2 \kappa \tanh \theta_s\, d\theta_s\, \Theta \Bigl[\cosh \theta_s - \cos(\kappa\sigma^2\Omega) \Bigr] (-\eta_0^2) \exp \left[-\frac{\log^2 \left(\frac {\cosh \theta_s} {b} \right)+(\theta_s-i\kappa \sigma^2\Omega)^2}{\kappa^2\sigma^2} \right] \frac {\kappa} {8\pi^2} \frac {\csch \theta_s} {\cosh \theta_{-s} - \cosh \theta_s}
\nonumber \\
&\quad = \frac {-\eta_0^2} {4\pi^2} \int_{\theta_s(0)}^{\theta_s(\infty) } d\theta\, s\, \frac {\Theta \Bigl[\cosh \theta - \cos(\kappa\sigma^2\Omega) \Bigr] } {b^2 - \cosh^2 \theta} \exp \left[-\frac{\log^2 \left(\frac {\cosh \theta} {b} \right)+(\theta-i\kappa \sigma^2\Omega)^2}{\kappa^2\sigma^2} \right] \,.
\end{align} 
\end{widetext}
A few things are worth noting here. First, while the limits of the integral are specified, the contour of integration is not a straight line. Instead it is some subset of a horseshoe-like curve similar to the one just above the real-$x$ axis in Fig.~\ref{fig:XPoles}. We will specify this shortly. Second, while we started with the change of variables $\theta_s(y)$ being dependent on~$s\in\{+,-\}$, by virtue of $\theta_s$ being promoted to an integration variable, it loses its $s$ dependence in that role (since any name for the integration variable will do), and so we rename it to just~$\theta$ in the second line. (Notice that the dependence on $\theta_{-s}$ has been eliminated since $\cosh \theta_s \cosh \theta_{-s} = b^2$.) The original integral itself is $s$~dependent, however, and this fact survives through the $s$-dependent contour of integration and the presence of an explicit $s$ in the integrand.

We will now specify the contour of integration. Plugging this integral back into Eq.~\eqref{eq:rescontrib1} gives
\begin{align}
\label{eq:rescontrib2}
	2\pi i & \sum_{\text{poles}} \binom {\text{residue}} {\text{integrals}} \nonumber \\ 
	&= \sum _{s \in \{+,-\}} \int_{\theta_s(0)}^{\theta_s(\infty) } d\theta\, s\, \Theta \Bigl[\cosh \theta - \cos(\kappa\sigma^2\Omega) \Bigr] I(\theta)\,,
\end{align}
where
\begin{align}
	I(\theta) = \frac {i \eta_0^2(2\pi)^{-1}} {\cosh^2 \theta - b^2} \exp \left[-\frac{\log^2 \left(\frac {\cosh \theta} {b} \right)+(\theta-i\kappa \sigma^2\Omega)^2}{\kappa^2\sigma^2} \right]\,.
\end{align}
The integration contour is not just a straight line. 
Nevertheless, we start with the endpoints of integration, which evaluate to
\begin{align}
	\theta_+(\infty) &
	= -\infty + i\pi\, \Theta[-b] \,, \\
	\theta_+(0) &= \arccosh(b)\,, \\
	\theta_-(\infty) &
	= \frac {i\pi} {2}\,, \\
	\theta_-(0) &= \arccosh(b)\,,
\end{align}
Since $y \ge 0$ and $b<1$, the ranges of $\theta_\pm(y)$ are respectively horseshoe- and L-shaped:
\begin{align}
	\theta_+ &\in (-\infty,0] \cup (0,i\pi) \cup [i\pi, - \infty + i\pi)\,, \\
	\theta_- &\in (0,i\pi) \cup [i\pi, \infty + i\pi)\,,
\end{align}
The Heaviside step function restricts these further, however, by requiring $\cosh \theta_\pm > \cos(\kappa \sigma^2 \Omega)$. Under this condition, the respective ranges become
\begin{align}
	\theta_+ &\in (-\infty,0] \cup (0,i\kappa \sigma^2 \Omega)\,, \\
	\theta_- &\in (0,i\kappa \sigma^2 \Omega)\,.
\end{align}
Note that these ranges will often be further restricted by the value of $b$, leaving $\theta_\pm(y)$ to explore only a subset thereof for all $y \ge 0$. The union of these total ranges (which accounts for any ${b<1}$) is shown as the thick L-shaped line between the raised integration contour and the real-$x$ axis in Fig.~\ref{fig:XPoles}. Using all this information, along with careful algebra, we can rewrite Eq.~\eqref{eq:rescontrib1} as
\begin{align}
\label{eq:rescontribfinal}
	&2\pi i \sum_{\text{poles}} \binom {\text{residue}} {\text{integrals}} \nonumber \\ 
	&= \begin{dcases}
\int_{i\, \text{Min}\left[\kappa\sigma^2\Omega,\frac{\pi}{2}\right]}^{i\kappa\sigma^2\Omega} d\theta\, I(\theta)\,, & b<0,\\
\left(\int_{i\, \text{Min}\left[\kappa\sigma^2\Omega,\frac{\pi}{2}\right]}^{0}+\int_{0}^{-\infty}\right) d\theta\, I(\theta)\,, & b>0.
\end{dcases}
\end{align}
The limits of integration provide two natural cases to consider: $b>0$ and $b<0$, which correspond to partially causal (Sec.~\ref{subsec:causalresidue}) and wholly noncausal (Sec.~\ref{subsec:noncausalresidue}) detectors, respectively. The integral for $b<0$ vanishes when $\kappa\sigma^2\Omega<\frac{\pi}{2}$ and otherwise is a finite integral along the imaginary-$\theta$ axis.  For $b>0$, the integral is an L-shaped contour integral along the positive imaginary-$\theta$ axis down to~0, then left along the negative real-$\theta$ axis from~0 to $-\infty$.

For the second case (${b>0}$)---in which the wedges defined by the detector trajectories overlap---the residue contribution can be calculated by directly applying the method of steepest descent to the contour integral, where the required saddle point is found numerically. We verified the validity of this approximation by comparing it to the result of numerically integrating along a modified contour that winds through the peaks and valleys in the complex-$\theta$ plane, passing through the saddle point
.

For the first case (${b<0}$)---spacelike separated detectors---the saddle point approximation succeeds (by the same test described above) when the imaginary part of the saddle point location is $< \kappa \sigma^2 \Omega$, but it fails otherwise. In the cases where it fails, we simply perform numerical integration to obtain the residue contribution.

Combining the two residue contributions with the residue-free portion yields the complete coherence term, $X$, by using Eq.~\eqref{eq:rescontribfinal} in Eq.~\eqref{eq:Xwithresidues}.

\subsection{Interpretation of residue contributions}
\label{subsec:residueinterp}

The second case in Eq.~\eqref{eq:rescontribfinal} ($b>0$) implies $L < 2\kappa^{-1}$, which corresponds to detector trajectories that are not fully causally separated---their respective Rindler wedges overlap. This case was discussed in Sec.~\ref{subsec:causalresidue}, wherein it was asserted that this contribution is due to the infinite tails of the Gaussian building up interactions along a light-like asymptote, which are eventually felt by the other detector (see Fig.~\ref{subsec:noncausalresidue}). This interpretation of the contribution is associated with the real part of the contour (from~0 to~$-\infty$), since $b>0$ and $y \to \infty$ implies $\theta_+ \to -\infty$.  Thus, the part of the contour running along the real-$x$ axis is associated with points along the two trajectories that are far separated in proper time, as is the case in Fig.~\ref{fig:ShockWave}.  These real-valued pole locations translate (via Eq.~\ref{eq:xytau}) to real values of $\tau$ and $\tau^\prime$ such that the corresponding points on the detector trajectories are null separated.

The first case ($b<0$) in Eq.~\eqref{eq:rescontribfinal} implies $L~>~2\kappa^{-1}$, which corresponds to detector trajectories that are causally disjoint (since their Rindler wedges have no intersection). This is the noncausal contribution discussed in Sec.~\ref{subsec:noncausalresidue}.  In this case, the pole locations are imaginary and correspond to points on the detector trajectories with $\Re(\tau)=-\Re(\tau^\prime)$.  However, $\tau$ and $\tau^\prime$ are both shifted in the imaginary direction by the same amount.

\subsection{Compact window functions}
\label{subsec:compactwindow}


In sections \ref{sec:resultsantipar} and \ref{sec:range} we described a number of features of entanglement harvesting using anti-parallel acceleration.  We discovered these features while carrying out the calculation of the negativity for anti-parallel detector systems.  The first two effects, enhancement over inertial detectors (Sec.~\ref{subsec:enhancement}) and entanglement resonance (Sec.~\ref{subsec:resonance}), arise in the residue-free part of the calculation. The second two features are causal (Sec.~\ref{subsec:causalresidue}) and noncausal (Sec.~\ref{subsec:noncausalresidue}) effects due to contributions from the residue terms. However, by choosing to use an analytic window function, we were afforded the ability to split the calculation into a residue-free contribution and a contribution from residue terms.  This ability is a mathematical artifact of our choice of analytic window function. This choice allows us to avoid direct numerical integration of highly oscillatory integrals, which proved to be excessively time consuming and was thus used only in limited circumstances (e.g., the entanglement resonance discussed in Sec.~\ref{subsec:resonance}). Throughout sections \ref{sec:resultsantipar} and \ref{sec:range} we make conjectures about changes in the negativity if we had used window functions that smoothly cut off to zero (see, e.g., Ref.~\cite{Garay:2014ef}). Such a compactly supported function, while smooth, cannot be analytic. Therefore, testing these conjectures would require completely different techniques, and we leave them to future work.

%
%
%
%
%

\bibliographystyle{bibstyleNCM_papers}
\bibliography{accel_ent-short,allrefs}

\end{document}